\def\eck#1{\left\lbrack #1 \right\rbrack}
\def\eckk#1{\bigl[ #1 \bigr]}
\def\rund#1{\left( #1 \right)}

\def\wave#1{\left\lbrace #1 \right\rbrace}
\def\ave#1{\left\langle #1 \right\rangle}

\def\part#1#2{{\partial #1\over\partial #2}}

\def\Re{{\cal R}\hbox{e}}

\def\A{{\cal A}}

\def\d{{\rm d}}

\def\eps{{\epsilon}}

\def\arcminf {\hbox{$.\!\!^{\prime}$}}
\def\vp{\varphi}
\def\vt{{\vartheta}}

\def\Real{{\rm I\mathchoice{\kern-0.70mm}{\kern-0.70mm}{\kern-0.65mm}%
  {\kern-0.50mm}R}}
\def\C{\rm C\kern-.42em\vrule width.03em height.58em depth-.02em
       \kern.4em}
\font \bolditalics = cmmib10
\def\bx#1{\leavevmode\thinspace\hbox{\vrule\vtop{\vbox{\hrule\kern1pt
        \hbox{\vphantom{\tt/}\thinspace{\bf#1}\thinspace}}
      \kern1pt\hrule}\vrule}\thinspace}

\def \vc #1{{\textfont1=\bolditalics \hbox{$\bf#1$}}}
{\catcode`\@=11
\gdef\SchlangeUnter#1#2{\lower2pt\vbox{\baselineskip 0pt \lineskip0pt
  \ialign{$\m@th#1\hfil##\hfil$\crcr#2\crcr\sim\crcr}}}
}
\def\gtrsim{\mathrel{\mathpalette\SchlangeUnter>}}
\def\lesssim{\mathrel{\mathpalette\SchlangeUnter<}}

\def\ueber#1#2{{\setbox0=\hbox{$#1$}%
  \setbox1=\hbox to\wd0{\hss$\scriptscriptstyle #2$\hss}%
  \offinterlineskip
  \vbox{\box1\kern0.4mm\box0}}{}}

\def\bx#1{\leavevmode\thinspace\hbox{\vrule\vtop{\vbox{\hrule\kern1pt
        \hbox{\vphantom{\tt/}\thinspace{\bf#1}\thinspace}}
      \kern1pt\hrule}\vrule}\thinspace}

 
\magnification=\magstep1
\input epsf
\voffset= 0.0 true cm
\vsize=19.8 cm     
\hsize=13.5 cm
\hfuzz=2pt
\tolerance=500
\abovedisplayskip=3 mm plus6pt minus 4pt
\belowdisplayskip=3 mm plus6pt minus 4pt
\abovedisplayshortskip=0mm plus6pt
\belowdisplayshortskip=2 mm plus4pt minus 4pt
\predisplaypenalty=0
\footline={\tenrm\ifodd\pageno\hfil\folio\else\folio\hfil\fi}

\def\la{\mathrel{\hbox{\rlap{\hbox{\lower4pt\hbox{$\sim$}}}\hbox{$<$}}}}
\def\ga{\mathrel{\hbox{\rlap{\hbox{\lower4pt\hbox{$\sim$}}}\hbox{$>$}}}}

\def\utw{\smash{\rlap{\lower5pt\hbox{$\sim$}}}}
\def\udtw{\smash{\rlap{\lower6pt\hbox{$\approx$}}}}

\def\getsto{\mathrel{\hbox{\rlap{$\gets$}\hbox{\raise2pt\hbox{$\to$}}}}}
\def\lid{\mathrel{\hbox{\rlap{\hbox{\lower4pt\hbox{$=$}}}\hbox{$<$}}}}
\def\gid{\mathrel{\hbox{\rlap{\hbox{\lower4pt\hbox{$=$}}}\hbox{$>$}}}}
\def\sol{\mathrel{\hbox{\rlap{\hbox{\raise4pt\hbox{$\sim$}}}\hbox{$<$}}}
}
\def\sog{\mathrel{\hbox{\rlap{\hbox{\raise4pt\hbox{$\sim$}}}\hbox{$>$}}}
}
\def\lse{\mathrel{\hbox{\rlap{\hbox{\raise4pt\hbox{$<$}}}\hbox{$\simeq$}
}}}
\def\gse{\mathrel{\hbox{\rlap{\hbox{\raise4pt\hbox{$>$}}}\hbox{$\simeq$}
}}}
\def\grole{\mathrel{\hbox{\lower2pt\hbox{$<$}}\kern-8pt
\hbox{\raise2pt\hbox{$>$}}}}
\def\leogr{\mathrel{\hbox{\lower2pt\hbox{$>$}}\kern-8pt
\hbox{\raise2pt\hbox{$<$}}}}
\def\loa{\mathrel{\hbox{\rlap{\hbox{\lower4pt\hbox{$\approx$}}}\hbox{$<$
}}}}
\def\goa{\mathrel{\hbox{\rlap{\hbox{\lower4pt\hbox{$\approx$}}}\hbox{$>$
}}}}

%
%

\font\kleinhalbcurs=cmmib10 scaled 833
\font\eightrm=cmr8
\font\sixrm=cmr6
\font\eighti=cmmi8
\font\sixi=cmmi6
\skewchar\eighti='177 \skewchar\sixi='177
\font\eightsy=cmsy8
\font\sixsy=cmsy6
\skewchar\eightsy='60 \skewchar\sixsy='60
\font\eightbf=cmbx8
\font\sixbf=cmbx6
\font\eighttt=cmtt8
\hyphenchar\eighttt=-1
\font\eightsl=cmsl8
\font\eightit=cmti8

\font\bxf=cmbx10
  \mathchardef\Gamma="0100
  \mathchardef\Delta="0101
  \mathchardef\Theta="0102
  \mathchardef\Lambda="0103
  \mathchardef\Xi="0104
  \mathchardef\Pi="0105
  \mathchardef\Sigma="0106
  \mathchardef\Upsilon="0107
  \mathchardef\Phi="0108
  \mathchardef\Psi="0109
  \mathchardef\Omega="010A
\def\rahmen#1{\vskip#1truecm}
\def\begfig#1cm#2\endfig{\par
\setbox1=\vbox{\rahmen{#1}#2}%
\dimen0=\ht1\advance\dimen0by\dp1\advance\dimen0by5\baselineskip
\advance\dimen0by0.4true cm
\ifdim\dimen0>\vsize\pageinsert\box1\vfill\endinsert
\else
\dimen0=\pagetotal\ifdim\dimen0<\pagegoal
\advance\dimen0by\ht1\advance\dimen0by\dp1\advance\dimen0by1.4true cm
\ifdim\dimen0>\vsize
\topinsert\box1\endinsert
\else\vskip1true cm\box1\vskip4true mm\fi
\else\vskip1true cm\box1\vskip4true mm\fi\fi}
\def\figure#1#2{\smallskip\setbox0=\vbox{\noindent\petit{\bf Fig.\ts#1.\
}\ignorespaces #2\smallskip
\count255=0\global\advance\count255by\prevgraf}%
\ifnum\count255>1\box0\else
\centerline{\petit{\bf Fig.\ts#1.\ }\ignorespaces#2}\smallskip\fi}

\def\xfigure#1#2#3#4{\midinsert\noindent
    $$\epsfxsize=#4truecm\epsffile{#3}$$
    \figure{#1}{#2}\endinsert}


\def\begtab#1cm#2\endtab{\par
\ifvoid\topins\midinsert\vbox{#2\rahmen{#1}}\endinsert
\else\topinsert\vbox{#2\kern#1true cm}\endinsert\fi}
\def\rahmen#1{\vskip#1truecm}
\def\begpet{\vskip6pt\bgroup\petit}
\def\endpet{\vskip6pt\egroup}

\def\begref{\par\bgroup\petit
\let\it=\rm\let\bf=\rm\let\sl=\rm\let\INS=N}
\def\petit{\def\rm{\fam0\eightrm}%
\textfont0=\eightrm \scriptfont0=\sixrm \scriptscriptfont0=\fiverm
 \textfont1=\eighti \scriptfont1=\sixi \scriptscriptfont1=\fivei
 \textfont2=\eightsy \scriptfont2=\sixsy \scriptscriptfont2=\fivesy
 \def\it{\fam\itfam\eightit}%
 \textfont\itfam=\eightit
 \def\sl{\fam\slfam\eightsl}%
 \textfont\slfam=\eightsl
 \def\bf{\fam\bffam\eightbf}%
 \textfont\bffam=\eightbf \scriptfont\bffam=\sixbf
 \scriptscriptfont\bffam=\fivebf
 \def\tt{\fam\ttfam\eighttt}%
 \textfont\ttfam=\eighttt
 \normalbaselineskip=9pt
 \setbox\strutbox=\hbox{\vrule height7pt depth2pt width0pt}%
 \normalbaselines\rm
\def\vec##1{\setbox0=\hbox{$##1$}\hbox{\hbox
to0pt{\copy0\hss}\kern0.45pt\box0}}}%
\let\ts=\thinspace
%
\font \tafontt=     cmbx10 scaled\magstep2
\font \tafonts=     cmbx7  scaled\magstep2
\font \tafontss=     cmbx5  scaled\magstep2
\font \tamt= cmmib10 scaled\magstep2
\font \tams= cmmib10 scaled\magstep1
\font \tamss= cmmib10
\font \tast= cmsy10 scaled\magstep2
\font \tass= cmsy7  scaled\magstep2
\font \tasss= cmsy5  scaled\magstep2
\font \tasyt= cmex10 scaled\magstep2
\font \tasys= cmex10 scaled\magstep1
\font \tbfontt=     cmbx10 scaled\magstep1
\font \tbfonts=     cmbx7  scaled\magstep1
\font \tbfontss=     cmbx5  scaled\magstep1
\font \tbst= cmsy10 scaled\magstep1
\font \tbss= cmsy7  scaled\magstep1
\font \tbsss= cmsy5  scaled\magstep1

\newbox\chsta\newbox\chstb\newbox\chstc
\def\centerpar#1{{\advance\hsize by-2\parindent
\rightskip=0pt plus 4em
\leftskip=0pt plus 4em
\parindent=0pt\setbox\chsta=\vbox{#1}%
\global\setbox\chstb=\vbox{\unvbox\chsta
\setbox\chstc=\lastbox
\line{\hfill\unhbox\chstc\unskip\unskip\unpenalty\hfill}}}%
\leftline{\kern\parindent\box\chstb}}
 \def \chap#1{
    \vskip24pt plus 6pt minus 4pt
    \bgroup
 \textfont0=\tafontt \scriptfont0=\tafonts \scriptscriptfont0=\tafontss
 \textfont1=\tamt \scriptfont1=\tams \scriptscriptfont1=\tamss
 \textfont2=\tast \scriptfont2=\tass \scriptscriptfont2=\tasss
 \textfont3=\tasyt \scriptfont3=\tasys \scriptscriptfont3=\tenex
     \baselineskip=18pt
     \lineskip=18pt
     \raggedright
     \pretolerance=10000
     \noindent
     \tafontt
     \ignorespaces#1\vskip7true mm plus6pt minus 4pt
     \egroup\noindent\ignorespaces}%
 \def \sec#1{
     \vskip25true pt plus4pt minus4pt
     \bgroup
 \textfont0=\tbfontt \scriptfont0=\tbfonts \scriptscriptfont0=\tbfontss
 \textfont1=\tams \scriptfont1=\tamss \scriptscriptfont1=\kleinhalbcurs
 \textfont2=\tbst \scriptfont2=\tbss \scriptscriptfont2=\tbsss
 \textfont3=\tasys \scriptfont3=\tenex \scriptscriptfont3=\tenex
     \baselineskip=16pt
     \lineskip=16pt
     \raggedright
     \pretolerance=10000
     \noindent
     \tbfontt
     \ignorespaces #1
     \vskip12true pt plus4pt minus4pt\egroup\noindent\ignorespaces}%
 \def \subs#1{
     \vskip15true pt plus 4pt minus4pt
     \bgroup
     \bxf
     \noindent
     \raggedright
     \pretolerance=10000
     \ignorespaces #1
     \vskip6true pt plus4pt minus4pt\egroup
     \noindent\ignorespaces}%
 \def \subsubs#1{
     \vskip15true pt plus 4pt minus 4pt
     \bgroup
     \bf
     \noindent
     \ignorespaces #1\unskip.\ \egroup
     \ignorespaces}
\def\footnoterule{\kern-3pt\hrule width 2true cm\kern2.6pt}
\newcount\footcount \footcount=0
\def\advftncnt{\advance\footcount by1\global\footcount=\footcount}
\def\fonote#1{\advftncnt$^{\the\footcount}$\begingroup\petit
       \def\textindent##1{\hang\noindent\hbox
       to\parindent{##1\hss}\ignorespaces}%
\vfootnote{$^{\the\footcount}$}{#1}\endgroup}

\newcount\sterne
\outer\def\byebye{\bigskip\typeset
\sterne=1\ifx\speciali\undefined\else
\bigskip Special caracters created by the author
\loop\smallskip\noindent special character No\number\sterne:
\csname special\romannumeral\sterne\endcsname
\advance\sterne by 1\global\sterne=\sterne
\ifnum\sterne<11\repeat\fi
\vfill\supereject\end}
\def\typeset{\centerline{\petit This article was processed by the author
using the \TeX\ Macropackage from Springer-Verlag.}}

\voffset=0pt

\chap{\centerline{Cluster mass profiles from weak lensing:} \hfill\break
\centerline{shear vs. magnification
information}}

\bigskip
\sec{\centerline{Peter Schneider, Lindsay King \& Thomas Erben}}
\bigskip
\noindent
\centerline{Max-Planck-Institut f\"ur Astrophysik} 
\centerline{Postfach 1523, D-85740 Garching, Germany}
\centerline{e-mail: peter, lindsay, erben@mpa-garching.mpg.de}

\sec{Abstract}

A massive foreground cluster lens changes the shapes (shear effect) 
and number density (magnification effect) of the faint
background galaxy population. In this paper we investigate how the
shear, magnification and
combined information can be used to constrain cluster mass profiles in
the weak lensing regime.
We develop maximum likelihood techniques to make quantitative
predictions for each of the methods. Our analytic results are checked
against Monte Carlo simulations.

In general, we find that the shear method is superior to the
magnification method. However, the magnification information 
complements the shear information if the former has accurate
external calibration. For the magnification method, we discuss the 
effects of random and systematic uncertainties in the background
galaxy counts. 

\sec{1 Introduction}
Weak gravitational lensing has been recognized as a powerful tool to
investigate the mass and mass distribution of clusters of galaxies
(e.g., Webster 1985; Tyson et al.\ 1990; Kochanek 1990). In their
pioneering paper, Kaiser \& Squires (1993) pointed out that the tidal
gravitational field of the cluster, as measured from the distortion of
image shapes of the faint background galaxy population, can be used to
reconstruct the two-dimensional projected mass profiles of clusters,
without referring to a parametrized mass model. This method, modified
in various ways later (e.g., Kaiser 1995; Schneider 1995; Seitz \&
Schneider 1995; Bartelmann et al.\ 1996; Seitz \& Schneider 1996;
Squires \& Kaiser 1996; Seitz \& Schneider 1997, 1998; Seitz,
Schneider \& Bartelmann 1998; Lombardi \& Bertin 1998a, 1998b) has
been applied to more than a dozen clusters up to now (e.g., Fahlman et
al.\ 1994; Smail et al.\ 1995; Squires et al.\ 1996a, b; Seitz et al.\
1996; Luppino \& Kaiser 1997; Clowe et al.\ 1998; Hoekstra et al.\
1998; Kaiser et al.\ 1999). One of the main results from these studies
is that in many clusters the projected mass distribution closely
follows the distribution of luminous cluster galaxies. For some of
these clusters, very large mass-to-light ratios have been reported,
e.g., for MS1224.7+2007, this ratio has been determined by two
independent studies to be larger than $\sim 800h$ in solar units
(Fahlman et al.\ 1994; Fischer 1999).

One of the main difficulties of these studies is the existence of the
so-called mass-sheet degeneracy (Falco et al.\ 1985; Schneider \&
Seitz 1995) which states that the image shapes are unchanged if the
surface mass distribution $\kappa(\vc\theta)$ (defined in the usual
way) is replaced by $\kappa(\vc\theta) \to \lambda \kappa(\vc\theta)
+(1-\lambda)$. In particular, this transformation keeps the critical
curves of the lens mapping fixed, and thus the location of the giant
arcs and arclets. The choice of the 
constant $\lambda$ in actual mass reconstructions is largely arbitrary.
Provided the observed data field is sufficiently large, one might
argue that the surface mass density has decreased to a value close to
zero far away from the cluster center, thereby fixing
$\lambda$. However, for those investigations where the available data
field is quite small, the invariance transformation implies a
substantial uncertainty in the mass estimates.

The mass sheet degeneracy affects the magnification as
$\mu(\vc\theta)\to \mu(\vc\theta)/\lambda^2$; hence, if magnification
information can be obtained, the degeneracy can be broken. Two
approaches to measuring the magnification have been presented in the
literature: Broadhurst et al.\ (1995) considered the effect of
magnification on the local number counts of faint background galaxies,
employing the so-called magnification bias effect. Since the number
counts of very faint galaxies are flatter than the critical slope
(defined below), magnification leads to a depletion of the local
number counts near the cluster center. This effect has been observed
in at least two clusters (Fort et al.\ 1997; Taylor et al.\
1998). Alternatively, Bartelmann \& Narayan (1995) have suggested
using the size of faint galaxies at fixed surface brightness as a
measure for the magnification, making use of the fact that
gravitational light deflection 
conserves the surface brightness. In addition, further measures of the
weak lensing magnification have been proposed, such as the change in
the redshift distribution of galaxies in a flux-limited sample
(Broadhurst et al.\ 1995), or the effect on the two-point angular
correlation function of distant galaxies (Moessner et al.\ 1998).

Whereas the assumption underlying the cluster mass reconstruction from
image ellipticities (the shear method, in what follows), namely that
the intrinsic orientation of source galaxies are randomly distributed
-- or that there is no preferred direction of alignment, is simple and
predicts that the mean ellipticity of galaxy images in a small region
of the cluster equals the reduced shear, the magnification techniques
need an external calibration: for the first of the methods mentioned
above, the number density of background galaxies in a `blank field'
as a function of limiting magnitude (and other selection criteria,
such as surface brightness etc.)
needs to be known, and for the
second, the mean size of galaxies as a function of surface brightness
is required. Uncertainties in this calibration information
immediately translates into uncertainties in the derived magnification. In
addition, naive estimates of the accuracy of shear and magnification
measurements (see Sect.\ 2) show that the shear method appears to be
superior.

Nevertheless, Broadhurst (1999) claimed that the magnification
information can, at least in some cases, yield a more accurate
determination of the projected radial mass profiles of clusters. In
this paper, we investigate the substance of this claim in more detail. In
Sect.\ 2 we briefly summarize the shear and magnification methods, and
provide a rough estimate of their respective accuracy in the
determination of local lens parameters. Sect.\ 3 describes a
maximum-likelihood method for the use of shear and magnification
information to determine mass profiles. The ensemble-averages of the 
likelihood functions are derived in Sect.\ 4; later, these are used to
determine the characteristic errors of parameter estimates.
In Sect.\ 5 we describe the numerical simulations used to substantiate
our analytic results, and the lens models that are considered.
We present the results from the likelihood analysis in Sect.\ 6; when
the unlensed background number density is known, we 
find that the shapes of the likelihood contours are different in both
methods. In particular, for a wide range of situations, the slope of
the mass profile is ill-determined from the shear method, and better
constrained by the magnification method. In Sect.\ 6 we also consider
the influence of uncertainty in the number density on the
likelihood analysis. We summarize our
results and conclude in Sect.\ 7.

\def\s{{\rm s}}
\def\L{{\cal L}}
\sec{2 Shear and magnification methods}
We use standard lensing notation; i.e., $\kappa(\vc\theta)$ denotes
the dimensionless surface mass density of the deflector -- which is
well defined if we assume that all source galaxies are at the same
distance, an assumption which is reasonable for low-redshift
clusters ($z_{\rm d}\lesssim 0.25$), and for many purposes not a bad
approximation even for medium-redshift lenses. The deflection potential
$\psi(\vc\theta)$ is related to $\kappa$ through a Poisson-like
equation, $\nabla^2\psi=2\kappa$. The trace-less part of the Hessian
of $\psi$ describes the tidal gravitational field, which we summarize
in the complex shear $\gamma=\gamma_1+{\rm i}\gamma_2
=(\psi_{,11}-\psi_{,22})/2+{\rm i}\psi_{,12}$, where indices separated
by a comma denote partial derivatives with respect to the position
$\vc\theta$ on the sky. The magnification of an image is the inverse
of the Jacobian determinant of the lens equation, $\mu(\vc\theta)=[\det
\A(\vc\theta)]^{-1}$, with $\det\A=(1-\kappa)^2-|\gamma|^2$. 

The shear method is based on the transformation between source and
image ellipticity due to the tidal field of the deflector. Let
$\eps^\s$ denote the intrinsic ellipticity of the source, and $\eps$
the ellipticity of the observed image. Here, we use the ellipticity
parameter as defined in, e.g.,  Seitz \& Schneider (1997), such that
$\eps$ is a complex number whose modulus, in the case of elliptical
isophotes with axis ratio $r$ is $|\eps|=(1-r)/(1+r)$, and whose phase
is twice the position angle of the major axis. The locally linearized
lens equation yields the transformation between source and image
ellipticity,
$$
\eps={\eps^\s + g\over 1+g^* \eps^\s}\; ,
\eqno (1)
$$
where $g=\gamma/(1-\kappa)$ is the complex reduced shear, and the
asterisk denotes complex conjugation. The
transformation (1) is valid only in the non-critical parts of the
lens, i.e., where $\det \A>0$, to which we will restrict the
discussion here. Let $p^\s_\eps(\eps^\s)\,\d^2\eps^\s$ be the probability
that the source ellipticity lies within $\d^2\eps^\s$ of $\eps^\s$,
then the expectation value of the image ellipticity is
$$
\ave{\eps}:=\int\d^2\eps^\s\;p^\s_\eps(\eps^\s)\,\eps(\eps^\s)=g\;,
\eqno (2)
$$
as shown by Schramm \& Kayser (1995) and Seitz \& Schneider
(1997). This result implies that the ellipticity of each galaxy image
provides an unbiased estimate of the local reduced shear (it should be
noted here that the reduced shear is unchanged under the mass-sheet
invariance transformation).

The magnification method based on the local number counts of
background galaxies arises from the magnification bias (e.g.,
Canizares 1982) which states that the local cumulative number counts
$n(\theta;S)$ above a flux limit $S$ are related to the unlensed counts
$n_0(S)$ by
$$
n(\theta;S)={1\over \mu}n_0\,\rund{S\over\mu}\; .
\eqno (3)
$$
If we assume that the number counts follow (locally) a power law of
the form $n_0\propto S^{-\beta}$, then
$$
n(\theta)=n_0\,\mu^{\beta-1}
\eqno (4)
$$
at any fixed flux threshold. This implies that if the intrinsic counts are
flatter than $1$, then the lensed counts will be reduced relative to
the unlensed ones. This number depletion is the signature of lensing
in the magnification method considered here.

Let us briefly consider the signal-to-noise of a shear and a
magnification detection by those respective methods. The noise in the
former case is due to the intrinsic ellipticity distribution, whereas
it is at least Poisson noise for the magnification method. If the
source galaxies have a significant angular correlation, the resulting
noise will be larger than the naive Poisson estimate used below;
we consider this in Sect.\ts 6.3.

Thus, assume that in some
(small) solid angle the shear is constant, and that there are
$N_\gamma$ background galaxy images for which the ellipticity can be
measured. The signal is $g$, and the noise is 
$\sigma_\eps/\sqrt{N_\gamma}$, where $\sigma_\eps$ is the ellipticity
dispersion. Thus, the signal-to-noise for a shear detection is
$$
\rund{\rm S\over N}_\gamma={|g|\sqrt{N_\gamma}\over \sigma_\eps}\;.
\eqno (5)
$$
Similarly, assume that the magnification is constant over a solid
angle and that, in the absence of lensing, $N_\mu$ galaxies would be
expected in that region. Then, the magnification signal is $|\Delta
N|=|\mu^{\beta-1} -1|N_\mu$, and the noise is $\sqrt{N_\mu}$,
ignoring the change of the number density in the noise estimate. Thus,
$$
\rund{\rm S\over N}_\mu=|\mu^{\beta-1} -1|\sqrt{N_\mu}\; .
\eqno (6)
$$
Specializing now to the case of weak lensing, i.e., $\kappa\ll 1$,
$|\gamma|\ll 1$, so that $\gamma\approx g$, 
the magnification becomes $\mu\approx 1+2\kappa$, and
a first-order expansion yields $|\Delta N|\approx
2\kappa|1-\beta|N_\mu$. Then, the ratio of the signal-to-noise
estimates becomes
$$
{({\rm S/ N})_\gamma\over ({\rm S/ N})_\mu}={|\gamma|\over \kappa}\,
{1\over 2\sigma_\eps |1-\beta|}\,\sqrt{N_\gamma\over N_\mu}\; .
\eqno (7)
$$
In a typical situation, the magnitudes of the shear and the surface
mass density are similar (they are exactly the same for a singular
isothermal sphere), and so the first factor in (7) is of order
unity. The slope of the galaxy number counts is flattest for the
faintest flux limits and for redder colors, and for these,
$\beta\gtrsim 0.5$. The variance of the intrinsic ellipticity is of
order $\sigma_\eps \approx 0.2$, and so the second factor in (7) is of
order 5. The last factor can be expected to be smaller than unity,
since the flux limit for counts can be fainter than that for which
ellipticity measurements are still possible, but in order to make up
for the second factor, the number density used for the counts would
need to be 25 times higher than for the shear estimates, which at the
assumed slope of the number counts would correspond to about 6
magnitudes! This of course is unrealistic -- the ratio of $N_\mu$ over
$N_\gamma$ is more typically of order 3 or 4 -- and we therefore
conclude from this simple analysis that the shear method is
considerably more sensitive than the magnification method based on
number counts. The other magnification method based on images sizes
yields a signal-to-noise ratio comparable to the shear method, but
since this effect has not been observed up to now, we shall not
consider this method in this paper.

\sec{3 Likelihood functions for shear and magnification}
We shall now consider the following situation: in a solid angle (e.g.,
around a cluster), one observes $N_\mu$ galaxy images at positions
$\vc\theta_i$, $1\le i\le N_\mu$, down to a given flux threshold. In
addition, one observes $N_\gamma$ galaxy images at positions
$\vc\vt_i$, $1\le i\le N_\gamma$, for which an ellipticity can be
measured; the two sets of galaxy images can have objects in common. If we now
assume that the mass distribution of the lens is described by a
parametrized model with parameters $\pi_i$, $1\le i\le M$, what are
the best-fitting parameters for the observation?

We solve this problem by defining a likelihood function and then
maximizing it. The likelihood is defined as the probability of
obtaining the observables given the model parameters $\pi_i$, and is
considered as the probability for the model parameters $\pi_i$ given the
observables (see, e.g., Press et al.\ 1992 for a discussion on this
point). 

Starting with the magnification effect, the probability of observing
$N_\mu$ galaxy images at positions $\vc\theta_i$ can be factorized in
the probability of observing a total of $N_\mu$ galaxies, and in the
probability that, given that there are $N_\mu$ galaxies, they are
located at $\vc\theta_i$. For a lens model with parameters $\pi_i$, the
expected number of galaxy images is
$$
\ave{N_\mu}=n_\mu\int\d^2\theta\; [\mu(\vc\theta)]^{\beta-1}\;,
\eqno (8)
$$
as follows from (4); we have used $n_\mu$ to denote the unlensed
number density, to distinguish it from the number density of galaxies
$n_\gamma$ for which an ellipticity can be measured, and the integral
extends over the solid angle of the observations. In practical
applications, this area can be quite complicated since accurate
photometry of very faint objects is impossible near brighter objects
-- like cluster galaxies -- which have to be masked out (see Taylor et
al.\ 1998 for an impressive illustration of this effect).  The
probability of finding $N_\mu$ galaxies in the data field follows a
Poisson distribution (if, as we shall assume, the source galaxies are
uncorrelated), and the probability that the $i$-th galaxy is located
at $\vc\theta_i$ is proportional to $n(\vc\theta_i)\propto
[\mu(\vc\theta_i)]^{\beta-1}$.  Then,
$$
\L_\mu=P(N_\mu;\ave{N_\mu})\prod_{i=1}^{N_\mu}
{[\mu(\vc\theta_i)]^{\beta-1}\over 
\int\d^2\theta\,[\mu(\vc\theta)]^{\beta-1}} \;,
\eqno (9)
$$
with
$$
P(N;\ave{N})={\ave{N}^N\over N!}\,\exp(-\ave{N})\; .
\eqno (10)
$$
One can now see that the denominator in the product on eq.\ (9) is
$\ave{N_\mu}/n_\mu$, and that therefore the factor
$\ave{N_\mu}^{N_\mu}$ of the Poisson distribution is cancelled. In
addition, the factors $N_\mu !$ and $n_\mu^{N_\mu}$ do not depend on
the lens model and yield just an irrelevant multiplicative constant in
the likelihood; therefore we shall drop this factor. Then, the
log-likelihood function becomes
$$
\ell_\mu := -\ln
\L_\mu=n_\mu\int\d^2\theta\;[\mu(\vc\theta)]^{\beta-1}
+(1-\beta)\sum_{i=1}^{N_\mu} \ln\mu(\vc\theta_i)\; .
\eqno (11)
$$
The best fitting model is now obtained by minimizing $\ell_\mu$ with
respect to the model parameters which are contained in the function
$\mu(\vc\theta)$. 

Next we turn to the likelihood function for the shear estimate. For
that, we need the probability that the observed ellipticity at
position $\vc\vt_i$ is $\eps$, given that the reduced shear is
$g(\vc\vt_i)$.  Assuming a Gaussian intrinsic ellipticity
distribution,
$$
p^\s_{\eps}(\eps^\s) = 
{\exp\rund{-|\eps^\s|^2/\sigma_\eps^2}
\over
\pi\sigma^{2}_\eps\left(1-\exp\rund{-1/\sigma_\eps^2}\right)}
\eqno (12)
$$
the observed probability distribution can be obtained from
$$
p_{\eps}(\eps |g) = p^\s_{\eps}(\eps^{\s}(\eps |g))\left
|{\partial^{2}\eps^{\s}}\over{\partial^{2}\eps}\right | = p^\s_{\eps}(\eps^{\s}(\eps |g))
{(\left | g\right
|^{2}-1)^{2} \over\left |\eps g^{*}-1\right |^{4}}\;,
$$
as in Geiger \& Schneider (1998). 
The corresponding probability density $p_\eps(\eps_i)$ for each lensed
galaxy in the catalog is determined, and the likelihood and
log-likelihood functions are given by
$$
\L_\gamma=\prod_{i=1}^{N_\gamma}{p_\eps(\eps_i|g(\vc\vt_i))}\;;\qquad
\ell_\gamma=-\sum_{i=1}^{N_\gamma}{\ln p_\eps(\eps_i|g(\vc\vt_i))}.\;
$$

We shall follow a somewhat simpler road in
the analytic treatment by assuming that the observed ellipticity 
distribution is a Gaussian;
this is a reasonable assumption if the intrinsic ellipticity
distribution is a Gaussian and if the reduced shear is not too close
to unity. The expectation value of $\eps$ is $g$, and the dispersion
in the two directions parallel and perpendicular to the shear are the
same (see Appendix of Geiger \& Schneider 1999) and given by
$$
\sigma^2=2\sigma_\bot^2=2\sigma_\Vert^2
=2\pi (1-|g|^2)^2\int_0^1 \d |\eps^\s|\;{|\eps^\s|^3\,
p^\s_\eps(|\eps^\s|)\over 1-|\eps^\s|^2 g^2}\;,
\eqno (13)
$$
as will be shown in the Appendix. Provided $|g|$ is much smaller than
unity, or that the intrinsic probability distribution does not extend
significantly out to $|\eps^\s|\sim 1$, the denominator in the
integral can be replaced by unity, for which the dispersion of the
observed ellipticities becomes
$$
\sigma \approx (1-|g|^2)\,\sigma_\eps\; .
\eqno (14)
$$
In the analytical treatment of Sect.\ts 4, we shall use the
approximation (13) for simplicity (for $\sigma_\eps =0.2$, the
approximation for $\sigma$ differs from the true value by 
less than 5\% for $|g|\leq$1, and less than 1\% for $|g|\leq$0.4). The likelihood function for the
shear method is then simply
$$
\L_\gamma=\prod_{i=1}^{N_\gamma}{1\over \pi\sigma^2[g(\vc\vt_i)]}
\exp\rund{-{|\eps_i-g(\vc\vt_i)|^2 \over \sigma^2[g(\vc\vt_i)]}} \;,
$$
and the corresponding log-likelihood function becomes, dropping
irrelevant additive constants,
$$
\ell_\gamma=\sum_{i=1}^{N_\gamma}\eck{
{|\eps_i-g(\vc\vt_i)|^2 \over \sigma^2[g(\vc\vt_i)]}
+2\ln\sigma[g(\vc\vt_i)] }\; .
\eqno (15)
$$
Again, the best-fitting lens model is obtained by minimizing
$\ell_\gamma$ with respect to the model parameters, which enter $g$
and thus also $\sigma$. 

The combined shear and magnification likelihood is then obtained by
just multiplying the respective likelihoods, or adding the
log-likelihoods, 
$$
\ell_{\rm tot}=\ell_\mu+\ell_\gamma\; .
\eqno (16)
$$

\def\t{{\rm t}}
\sec{4 Ensemble-averaged log-likelihood}
For a given set of observables, the likelihood functions defined in
the previous section can be calculated; the minimum yields the
best-fitting model parameters, and the width of the likelihood
function yields the confidence region for the parameter estimate from
this data set. In order to determine the characteristic errors of
parameter estimates, we consider here the ensemble-averaged
log-likelihood function.

Given a lens model -- hereafter called the true model, and
characterized by subscripts `t' -- the ensemble-average of a quantity
$X$ is provided by
$$
\ave{X}:=\sum_{N=0}^\infty P(N;\ave{N}_\t)
\eck{\prod_{i=1}^N\int\d^2\theta_i\;p_\t(\vc\theta_i)
\int\d^2\eps_i\;p_\t(\eps_i)} \; X\; ,
\eqno (17)
$$
where the first term averages over the probability to find $N$
galaxies when the expectation value from the true model is
$\ave{N}_\t$, the second term is the integration over the probability
that galaxy number $i$ lies at $\vc\theta_i$, and the final term
integrates over the ellipticity distribution. To see how this
procedure works, we start with the log-likelihood function for the
magnification method, i.e., setting $X=\ell_\mu$. Since $\ell_\mu$
does not depend on the ellipticity of the images, the final
integration in (17) yields simply a factor one. The probability for
the positions of the galaxies is
$$
p_\t(\vc\theta_i)={[\mu_\t(\vc\theta_i)]^{\beta-1}\over 
\int\d^2\theta\, [\mu_\t(\vc\theta)]^{\beta-1}}\; .
\eqno (18)
$$
Then,
$$
\ave{\ell_\mu}=\sum_{N_\mu=0}^\infty P(N_\mu;\ave{N_\mu}_\t)
\eck{\prod_{i=1}^{N_\mu}\int\d^2\theta_i\;p_\t(\vc\theta_i)}\,\ell_\mu\;,
\eqno (19)
$$
with 
$$
\ave{N_\mu}_\t=n_\mu\int\d^2\theta\;[\mu_t(\vc\theta)]^{\beta-1}\; .
\eqno (20)
$$
The first term in the log-likelihood function (11) depends neither on
$\vc\theta_i$ nor on the actual number $N_\mu$ of galaxies, and thus
the averaging operator in (19) leaves this term unchanged. The second
term of $\ell_\mu$ consists of a sum over individual galaxy images;
therefore, for the $j$-th galaxy image, only the term $i=j$ in the
spatial averaging in (19) contributes, the other integrations just
yield a factor one. Since there are $N_\mu$ equal terms, one obtains
$$\eqalign{
\ave{\ell_\mu}&=n_\mu\int\d^2\theta\;[\mu(\vc\theta)]^{\beta-1} \cr
&+(1-\beta)\sum_{N_\mu=0}^\infty P(N_\mu;\ave{N_\mu}_\t)
{N_\mu\over \int\d^2\theta\, [\mu_\t(\vc\theta)]^{\beta-1}}
\int\d^2\theta\;[\mu_\t(\vc\theta)]^{\beta-1}\,\ln\mu(\vc\theta)
\; . \cr }
\eqno (21)
$$
The sum over $N_\mu$ can now be performed, and using (20) we
obtain
$$
\ave{\ell_\mu}=n_\mu\int\d^2\theta\;[\mu(\vc\theta)]^{\beta-1}
+n_\mu(1-\beta)\int\d^2\theta\;
[\mu_\t(\vc\theta)]^{\beta-1}\,\ln\mu(\vc\theta) \;.
\eqno (22)
$$
From the derivative of $\ave{\ell_\mu}$ with respect to the parameter
$\pi_i$, 
$$
{\partial\ave{\ell_\mu}\over\partial \pi_i}
=(1-\beta)n_\mu \int\d^2\theta\;{1\over \mu(\vc\theta)}\,
{\partial \mu(\vc\theta)\over\partial
\pi_i}\,\wave{[\mu_\t(\vc\theta)]^{\beta-1}
-[\mu(\vc\theta)]^{\beta-1}}\;,
\eqno (23)
$$
one immediately sees that $\ave{\ell_\mu}$ attains a minimum when the
$\pi_i$ attain their true values, i.e., when $\mu=\mu_\t$. 

If one assumes that the likelihood function behaves approximately like
a Gaussian near its maximum, then the second partial derivatives of
$\ave{\ell_\mu}$ contain the information about the confidence region,
and so we define
$$
V_{ij}^\mu:={\partial^2\ave{\ell_\mu}\over
\partial \pi_i\partial \pi_j}_{|\pi=\pi_\t}
=(1-\beta)^2 n_\mu \int\d^2\theta\,[\mu_\t(\vc\theta)]^{\beta-3}
\rund{ {\partial\mu(\vc\theta)\over \partial \pi_i}
{\partial\mu(\vc\theta)\over \partial \pi_j}}_{|\pi=\pi_\t}\; ,
\eqno (24)
$$
where the values of the $\pi$ derivatives are evaluated at the true
parameter values. 

Next, we consider the ensemble average of the log-likelihood function
for the shear. For that we have to specify the probability
distribution for the image ellipticities for the true model. As
mentioned before, we shall use a Gaussian with mean $g_\t(\vc\vt)$,
the true reduced shear at position $\vc\vt$, and dispersion given by
(13), again with $g=g_\t$, and denoted by $\sigma_\t$. Then, from (15)
and (17), we see that, since $\ell_\gamma$ is a sum over galaxy
images, the integration operators work term by term, and so we get
$N_\gamma$ identical terms from the integrations. This then
immediately allows one to perform the sum over $N_\gamma$, as before, so that
$$\eqalign{
\ave{\ell_\gamma}&=\ave{N_\gamma}_\t\int\d^2\vt\;
{[\mu_\t(\vc\vt)]^{\beta-1}\over
\int\d^2\theta\, [\mu_\t(\vc\theta)]^{\beta-1}} \cr &\times
\int\d^2\eps {1\over \pi \sigma_\t^2(\vc\vt)}\,
\exp\rund{-{|\eps-g_\t(\vc\vt)|^2\over \sigma_\t^2(\vc\vt)}}\,
\rund{{|\eps-g(\vc\vt)|^2\over \sigma^2(\vc\vt)}+2\ln\sigma(\vc\vt)}\;
,\cr }
\eqno (25)
$$
and $\ave{N_\gamma}_\t$ is defined in analogy to (20), with $n_\mu$
replaced by $n_\gamma$. The $\eps$-integration is readily performed,
and one obtains
$$
\ave{\ell_\gamma}=n_\gamma\int\d^2\vt\;[\mu_\t(\vc\vt)]^{\beta-1}
\rund{{|g(\vc\vt)-g_\t(\vc\vt)|^2+\sigma_\t^2(\vc\vt)\over 
\sigma^2(\vc\vt)}+2\ln\sigma(\vc\vt)}\; .
\eqno (26)
$$
By taking a first partial derivative of $\ave{\ell_\gamma}$ with
respect to the parameter $\pi_i$, 
$$
{\partial\ave{\ell_\gamma}\over \partial \pi_i}
=n_\gamma\int\d^2\vt\;[\mu_\t(\vc\vt)]^{\beta-1}\;{\partial Y\over
\partial \pi_i}\;,
\eqno (27)
$$
with 
$$
{\partial Y\over \partial \pi_i} ={2\over \sigma^2}\Re\eck{{\partial
g\over \partial \pi_i} (g^*-g_\t^*)}
-{2\eckk{|g-g_\t|^2+(\sigma_\t^2-\sigma^2)}\over \sigma^3}
{\partial \sigma\over \partial \pi_i} \; ,
\eqno (28)
$$
we see that $\ave{\ell_\gamma}$ takes a minimum if $\pi=\pi_\t$, so that
$g=g_\t$ and therefore $\sigma=\sigma_\t$. 

From the ensemble-averaged log-likelihood function, we estimate the
dispersion of the parameters one obtains from a single realization of
the observables by using the fact that asymptotically (in the limit of
infinitely many data points) the distribution of $2\Delta\ell$ tends to
a $\chi^2_M$ distribution with $M$ being the number of model
parameters. Since the number of galaxies is finite, the likelihood
function is not Gaussian, and this estimate can only be 
approximate. We shall demonstrate with simulations that it provides
a very good approximation indeed.

\sec{5 Simulations and models}
The situation we shall consider is that of an axially-symmetric mass
distribution for the cluster. We assume that the central part of the
cluster is a strong lens where arcs and multiple images are
found, and from their modelling the Einstein radius $\theta_{\rm E}$
of the cluster is known. We then assume that the weak lensing analysis
is performed in an annulus with inner and outer radii $\theta_{\rm
in}$ and $\theta_{\rm out}$, with $\theta_{\rm in} > \theta_{\rm
E}$. If $\beta<1$ (we shall assume throughout the paper that
$\beta=0.5$), the number density of galaxies in the annulus is always
smaller than the unlensed number density $n_0$, since $\mu\ge 1$ outside
the critical curve (magnification theorem, see Schneider 1984). Hence,
we can use the rejection method, as explained below.

\subs{5.1 Description of the simulations}
Given a specific lens model, galaxy images are distributed in the
following way: First, the expected number of galaxies in the annulus
in the absence of lensing is calculated, $\ave{N_0}=n_0\pi(\theta_{\rm
out}^2-\theta_{\rm in}^2)$, and a number $N_0$ is drawn from a Poisson
distribution with mean $\ave{N_0}$. Then, these $N_0$ galaxies are
provisionally distributed in angular radius, by drawing a random
number $\xi$ uniformly distributed in $[0,1]$, and assigning a radius
of $\theta=\sqrt{\xi(\theta_{\rm out}^2-\theta_{\rm in}^2)+\theta_{\rm
in}^2}$. Then, a second uniform deviate $\eta\in[0,1]$ is drawn, and a
galaxy at $\theta$ is put into the final `catalog' only if
$[\mu(\theta)]^{\beta-1}\ge \eta$; otherwise it is discarded. With
this rejection method, the distribution of galaxies follows the
distribution (18). For each galaxy in the `catalog', a source
ellipticity $\eps^\s$ is drawn from a two-dimensional Gaussian
probability density distribution given by equation (12), and the
corresponding image ellipticity is obtained from (1). As a result, we
obtain a catalog containing radial positions $\theta_i$ and image
ellipticities $\eps_i$. For such a catalog, and an appropriately
parametrized family of lens models, the log-likelihood functions
$\ell_\mu$, $\ell_\gamma$ and $\ell_{\rm tot}$ can be minimized,
resulting in the best-fitting parameters.

The number density of galaxies for the shear and the magnification
method can be different; throughout we assume that $n_\gamma=30{\rm
arcmin}^{-2}$, and $n_\mu=120{\rm arcmin}^{-2}$ (the former of these
values is the typical number density used for shear analysis from
deep ground-based data, the latter is similar to those used by Fort et
al.\ 1997 and by Broadhurst 1999). Furthermore, we shall
assume that the Einstein radius of the cluster is $\theta_{\rm
E}=0.5{\rm arcmin}$, as is approximately the case for the cluster
CL0024+17 to which the magnification method has been applied (Fort et
al.\ 1997; Broadhurst 1999). The resulting confidence regions of the
parameter estimates depend solely on the ratios $\theta_{\rm
in}/\theta_{\rm E}$, $\theta_{\rm out}/\theta_{\rm E}$, and on the
products $\theta_E^2n_\gamma$ and $\theta_E^2 n_\mu$, and so can be
easily scaled for different values of the Einstein radius.

\subs{5.2 Lens models}
We shall consider two families of two-parameter lens models. The first
(Family A)
is characterized by a power-law profile outside the Einstein radius,
i.e., 
$$
\kappa(\theta)= a\rund{\theta\over \theta_{\rm E}}^{-q}\; {\rm for}\;
\theta\ge \theta_{\rm E}\; ;
\eqno(29)
$$
hence, $a$ is the surface mass density at the Einstein radius. For
example, a singular isothermal sphere would have $a=0.5$ and
$q=1$. Note that we do not have to specify the mass distribution
inside the Einstein radius; it suffices to know that the mean surface
mass density inside the Einstein radius is $\bar \kappa=1$. Using the
methods described in Chap.\ 8 of Schneider et al.\ (1992; hereafter
SEF), we find that
$$
\bar\kappa(\theta)=\rund{1-{2a\over 2-q}}
\rund{\theta\over \theta_{\rm E}}^{-2}
+{2a\over 2-q}\rund{\theta\over \theta_{\rm E}}^{-q}\; ,
\eqno (30)
$$
and 
$$
\gamma(\theta)=\bar\kappa(\theta)-\kappa(\theta)\; .
\eqno (31)
$$
Thus, this family has a `shape' parameter $q$ and a `strength' parameter
$a$. 

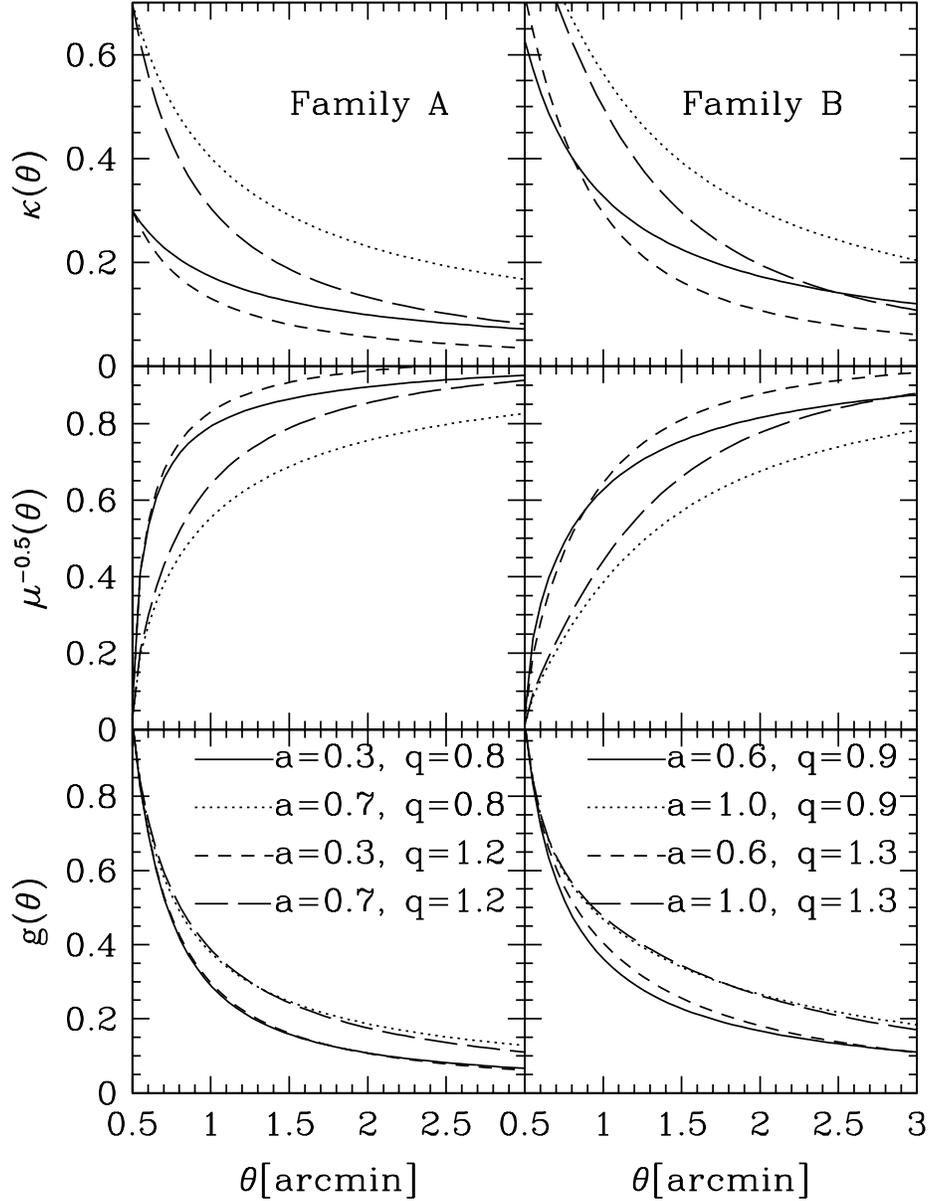
\xfigure{1}{The radial mass profile $\kappa$ (upper panels), the
magnification signature for the image density depletion, $\mu^{-0.5}$
(middle panels), and the reduced shear $g$ (bottom panels) as a
function of angular separation from the cluster center, for four
different combinations of the model parameters $a$ and $q$, as
indicated. Left panels are for models of Family A, right panels for
Family B}{fig1.tps}{13}

For the second family of lens models (hereafter Family B), we use the
model described in Sect.\ 8.1.5 of SEF, which reads
$$
\kappa(\theta)=\kappa_0{1+p(\theta/\theta_{\rm c})^2\over 
[1+(\theta/\theta_{\rm c})^2]^{2-p}}\; ,
\eqno (32)
$$
where $\theta_{\rm c}$ is the core radius, and $\kappa_0$ the central
surface mass density. For $\theta\ll \theta_{\rm c}$, the surface mass
density is nearly constant at $\kappa_0$, whereas for $\theta\gg
\theta_{\rm c}$ it decreases like a power law, with a slope determined
by $p$. The Einstein radius is given as
$$
\theta_{\rm E}=\theta_{\rm c}\sqrt{\kappa_0^{1/(1-p)}-1}=:
\theta_{\rm c} \sqrt{W}\;,
\eqno (33)
$$ 
and only exists if $\kappa_0>1$, a well-known result for
centrally-concentrated axially-symmetric lenses. Hence, we can
eliminate $\theta_{\rm c}$ in terms of $\theta_{\rm E}$ in (32) and
again obtain a two-parameter model (for fixed $\theta_{\rm E}$),
$$
\kappa(\theta)=\kappa_0\,{1+pW(\theta/\theta_{\rm E})^2\over 
[1+W(\theta/\theta_{\rm E})^2]^{2-p}}\; ,
\eqno (34)
$$
which leaves $\kappa_0$ and $p$ as the two parameters. Asymptotically,
for $\theta\to \infty$, the mass distribution behaves like
$$
\kappa\to \kappa_0\, p\, W^{p-1} (\theta/\theta_{\rm E})^{2p -2}\;.
$$
We shall choose the two free parameters in such a way that the
asymptotic form of this model is similar to that of Family A. Hence we
take $a=\kappa_0 p W^{p-1}$ and $q=2-2p$ as our parameters. Then,
$1/W=[2a/(2-q)]^{2/q}-1$, $\kappa_0=(1+W)^{q/2}$, and therefore,
$$
\kappa(\theta)=\rund{2a\over 2-q}{
\rund{2a\over 2-q}^{2/q}-1+\rund{2-q\over 2}
\rund{\theta\over\theta_{\rm E}}^2\over 
\eck{\rund{2a\over 2-q}^{2/q}-1 + \rund{\theta\over\theta_{\rm E}}^2}
^{1+q/2}}\; ;
\eqno (35)
$$
accordingly,
$$
\bar\kappa(\theta)=\rund{2a\over 2-q}
\eck{\rund{2a\over 2-q}^{2/q}-1+\rund{\theta\over\theta_{\rm
E}}^2}^{-q/2}\; .
\eqno (36)
$$
As before the shear is calculated from (31). The condition
$\kappa_0>1$, or $W>0$,
translates into
$$
a>{2-q\over 2}\; .
\eqno (37)
$$
For $a=(2-q)/2$, the resulting lens model coincides with the
corresponding one from Family A.

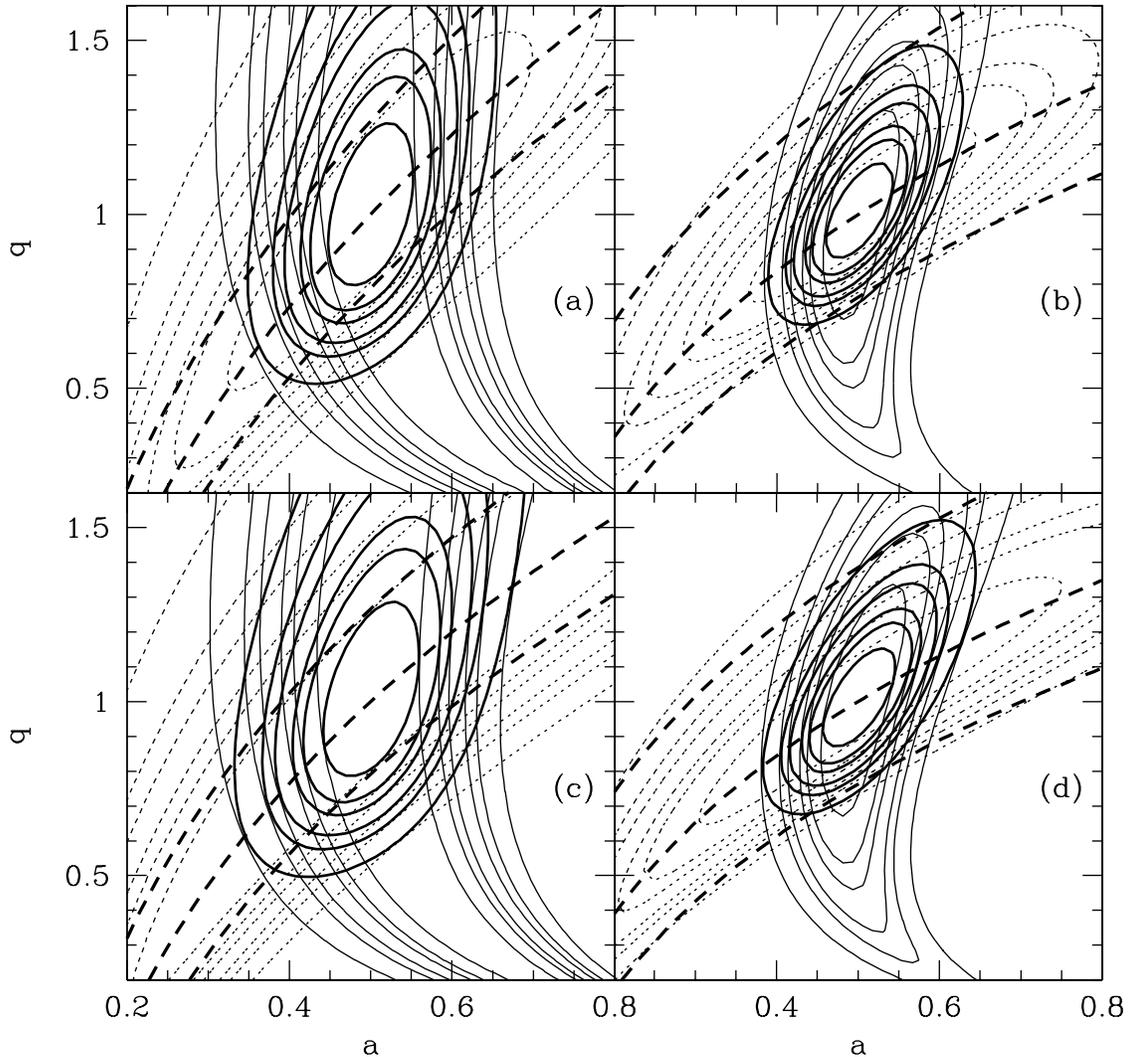
\xfigure{2}{Contours of constant average likelihood for models of Family
A. The thin solid lines are contours of constant
$\ave{\ell_\gamma}$, the dashed contours correspond to
$\ave{\ell_\mu}$, and the heavy solid contours to $\ave{\ell_{\rm
tot}}$. Contours are drawn for $2 \Delta\ell=\{2.30, 4.61, 6.17, 9.21,
11.8, 18.4\}$, within which one expects that 68.3\%, 90\%, 95.4\%, 99\%,
99.73\%, and 99.99\% respectively, of parameter estimates from
realizations will be enclosed. The input model is described by
$a=0.5$, $q=1.0$, and corresponds to the minimum of
$\ave{\ell}$. Parameters are as described in the text, i.e.,
$n_\mu=120\,{\rm arcmin}^{-2}$, $n_\gamma=30\,{\rm arcmin}^{-2}$,
$\beta=0.5$, $\sigma_\eps=0.2$, and the inner and outer radii of the
annulus are (a) $\theta_{\rm in}=0\arcminf 6$, $\theta_{\rm
out}=2\arcminf 0$; (b) $\theta_{\rm in}=0\arcminf 6$, $\theta_{\rm
out}=4\arcminf 0$; (c) $\theta_{\rm in}=0\arcminf 9$, $\theta_{\rm
out}=2\arcminf 0$; (d) $\theta_{\rm in}=0\arcminf 9$, $\theta_{\rm
out}=4\arcminf 0$.  The three heavy solid curves are curves of
constant $\ave{N_\mu}$; the middle one is for $\ave{N_\mu}$ fixed at
the value obtained for the input model, the two others are for changes
of $\ave{N_\mu}$ by $\pm 5\%$ }{fig2.tps}{15}
In Fig.\ 1 we plot, for four different combinations of $a$ and $q$,
the radial mass profile $\kappa(\theta)$, the magnification signature
$\mu^{\beta-1}$ for $\beta=0.5$, and the reduced shear $g$, for both
Families of models. The upper panels show the radial dependence of
the surface mass density $\kappa(\theta)$ which are all fairly
different over the range of angles plotted. The bottom panels show the
radial dependence of the reduced shear; here we see, in particular for
Family A, that models with the same amplitude but different slope have
nearly the same curves $g(\theta)$; hence, it will be quite
challenging to determine the radial slope $q$ for these models from the
shear method. The reason for this similarity comes from an accidental
cancellation of the slope dependencies of $1-\kappa$ and $\gamma$
in $g$ for these models. In contrast to the reduced shear,
the magnification signal, plotted as $\mu^{\beta-1}=\mu^{-0.5}$ in the
middle panels, differs clearly between models with different
slopes. Hence, despite the conclusion reached in Sect.\ts 2 on the
relative sensitivity of the shear and magnification methods, from the
curves in Fig.\ts 1 one might expect that the magnification effect could
more easily distinguish between different radial slopes.

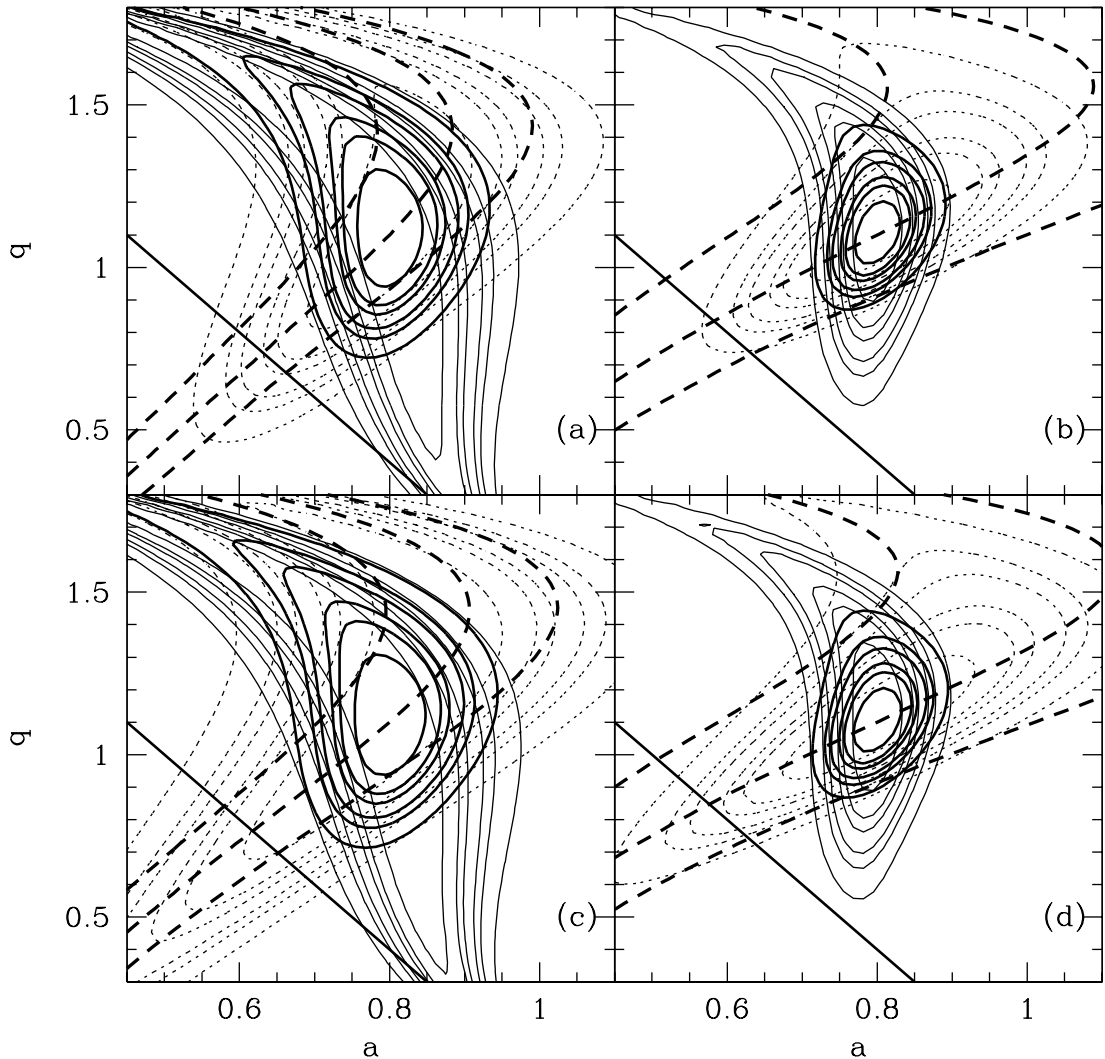
\xfigure{3}{Same as Fig.\ts 2, but now for models of Family B. The
input model is characterized by $a=0.8$, $q=1.1$. Inner and outer
radii are the same as in Fig.\ts 2. The heavy solid line in each panel
denotes the limit in (37); models below this line do not correspond to
centrally condensed mass distributions with $\kappa_0>1$}{fig3.tps}{15}

\sec{6 Results from the likelihood analysis}
We now proceed to consider the likelihood functions for the models
described in the previous Section, starting with the analytic 
results in Sect.\ts 6.1. In Sect.\ts 6.2 the results of the numerical
simulations are used to substantiate our analytic treatment. 
We also ask whether it is possible to distinguish between
lens Families, using each of the methods. In Sect.\ts 6.3 we consider 
how the uncertainty in the unlensed background number density
influences the magnification likelihood functions.

\subs{6.1 Analytic results}
In Figs.\ts 2 and 3, contours of
constant likelihood are plotted, for $\ave{\ell_\gamma}$,
$\ave{\ell_\mu}$, and $\ave{\ell_{\rm tot}}$, for models of Family A
and B, respectively. The contours are such that one expects the
parameters obtained by minimizing the log-likelihood from realizations
of galaxy image observations to lie within the contours in 68.3\%,
90\%, 95.4\%, 99\%, 99.73\%, and 99.99\% percent of all cases. This
expectation is based on the approximation mentioned at the end of
Sect.\ts 4. The various panels in the two figures differ in the choice
of the inner and outer radii of the annulus in which the data are
assumed to be analyzed.

Considering Fig.\ts 2 first, the first thing to note is that the
contours obtained from the shear method are very nearly parallel to
the $q$-axis. This is in accordance with our discussion of Fig.\ts 1
at the end of the previous section where we have seen that the reduced
shear depends only weakly on the slope $q$ of the mass
profile. Secondly, the direction of elongation of the likelihood
contours corresponding to the magnification method is nicely off-set
from the $q$-axis; this implies that the combination of shear and
magnification information is much more powerful than each method
individually in distinguishing between the model parameters. This fact
is actually demonstrated by the likelihood contours corresponding to
$\ave{\ell_{\rm tot}}$ which are much more compact than the other two
sets of contours. Considering the upper left panel as an example,
the 1-$\sigma$ contour of the magnification method
closes within the boundary of the figure, whereas the 1-$\sigma$
contour of $\ave{\ell_\gamma}$ is open both to the top and the bottom;
hence, from the magnification alone one can constrain the slope and
the amplitude, whereas the shear method yields basically no constraint
on the slope. Increasing the outer radius of the annulus, the shear
method becomes substantially more sensitive; in the upper right panel
of Fig.\ts 2, the areas enclosed by the corresponding contours for the
shear and the magnification methods are about the same.

In order to check whether most of the information comes from galaxy
images very close to the critical curves at $\theta=\theta_{\rm
E}=0\arcminf 5$, we have increased the inner radius of the annulus
in the lower two panels. Although the likelihood contours widen
slightly, this is not a dramatic effect; in particular, the 
likelihood contours for the combination of both methods remain
basically unchanged. Therefore, most of the signal is not located
right at the Einstein radius. This is good news, since there the
measurement of galaxy shapes and fluxes may be affected by bright
galaxies in the center of the cluster.

For the models of Family B, the situation is qualitatively the same as
discussed for Family A. Here, the advantage of increasing the outer
radius of the annulus is even larger for the shear method than was the
case for the Family A models. This result shows the importance of
wide-field images for a quantitative analysis of cluster mass profiles
(see Bonnet et al.\ 1994). 
As for Family A, little information is lost by
increasing the inner radius from $1.2\theta_{\rm E}$ to
$1.8\theta_{\rm E}$. 

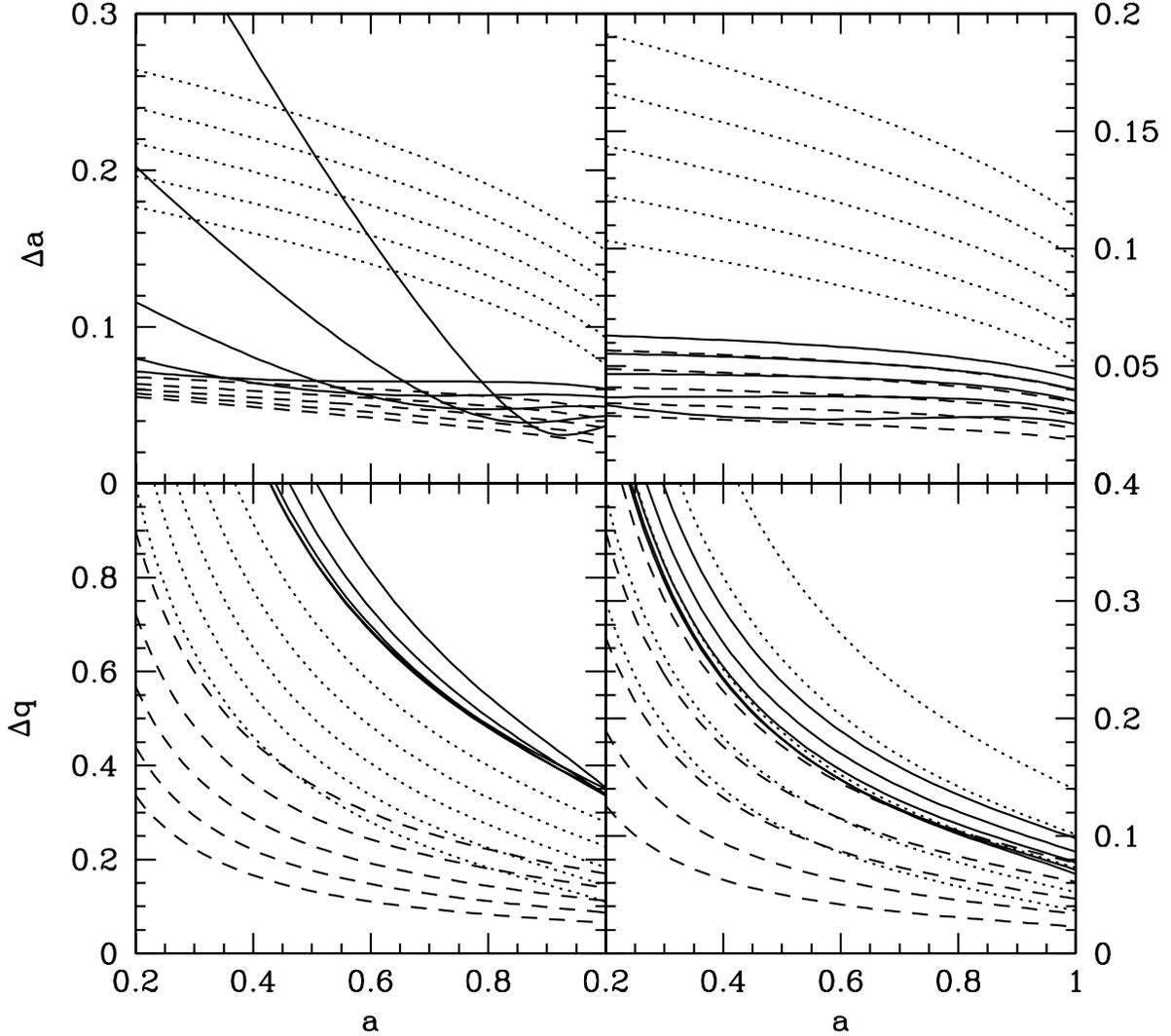
\xfigure{4}{The errors $\Delta a$ and $\Delta q$, as defined in the
text, plotted as a function of the input value of $a$. The lens model
Family A is considered here. The left panels
are for $\theta_{\rm in}=0\arcminf 6$, $\theta_{\rm out}=2\arcminf 0$,
and the right panels for $\theta_{\rm in}=0\arcminf 6$, $\theta_{\rm
out}=6\arcminf 0$.  Each panel shows three sets of five curves; the
solid curves correspond to the shear method, i.e., are based on
$\ave{\ell_\gamma}$, the dotted curves correspond to the magnification
method, and the dashed curves are the errors from using the
combination of shear and magnification information. Curves within each
set correspond to different input values of $q$, with $q=0.6, 0.8,
1.0, 1.2, 1.4$.  For $\ave{\ell_\mu}$ and $\ave{\ell_{\rm tot}}$, the
curves vary monotonically with $q$, the lowest corresponding to
$q=0.6$, the highest to $q=1.4$. The dependence of the errors for
$\ave{\ell_\gamma}$ on $q$ is more complicated and not necessarily
monotonic; for small $a$, the uppermost curves in the left panels
correspond to $q=0.6$, for the upper right panel, the behaviour is
monotonic, with $\Delta a$ increasing with $q$. For the lower
right panel, the two largest values of $\Delta q$ correspond to the
two largest values of $q$; the curves for the other values of $q$ are 
almost coincident}{fig4.tps}{16}

In order to investigate a larger part of parameter space, we consider
the behaviour of the log-likelihood function near the minimum. Writing
$$
\ave{\ell}(\pi)\approx\ave{\ell}(\pi_\t)+{1\over 2}{\partial^2 \ave{\ell}\over
\partial \pi_i\,\partial \pi_j}\,(\Delta \pi)_i\,(\Delta \pi)_j
\eqno (38)
$$
for parameter values close to those of the `true' parameters, curves
of constant $\Delta \ave{\ell}$ are described as ellipses. The
`1-$\sigma$ error ellipse' is described by setting $2\Delta
\ave{\ell}=2.30\equiv X$. Note that from Figs.\ts 2 and 3 one finds that the
confidence contours are not always well approximated by ellipses;
nevertheless, the consideration here should provide a qualitative
handle on the expected uncertainties in parameter determinations. The
matrix of partial derivatives in (38) has been calculated for the
magnification method in (24); the corresponding expression for the
shear method is too long to be reproduced here. The  `1-$\sigma$ error
ellipse' is characterized by three numbers, corresponding to the major
and minor axes, and their orientation. In order to translate this into
errors $\Delta a$ and $\Delta q$ for parameter estimates, we chose the
following prescription: We construct the smallest rectangle which
encloses the ellipse, and take as errors the half-lengths of its
sides. Denoting the matrix in (38) by $V$, one then obtains
$$\eqalign{
{\Delta a\over \sqrt{X}}& =\sqrt{\det V\over V_{11}^2 V_{22}}
+{V_{12}^2\over V_{11}}\sqrt{1\over V_{22}\det V}\; ,\cr
{\Delta q\over \sqrt{X}}& =\sqrt{\det V\over V_{11} V_{22}^2}
+{V_{12}^2\over V_{22}}\sqrt{1\over V_{11}\det V}\; .\cr }
\eqno (39)
$$
This graphical prescription for the error in the two parameters is
certainly not optimal, but easy to apply; in cases where the
eigendirections of $V$ are close to `diagonal' (as is the case for the
magnification method in Fig.\ts 2), this method can severely overestimate
the error.

The errors determined in that way are plotted in Fig.\ts 4, for two
choices of the radii of the annulus in which observations are assumed
to be available; the small (large) outer radius corresponds to the left
(right) panels in Fig.\ts 4. They are plotted as a function of the
value of $a$, for five different values of $q$, and using the
shear (solid curves), magnification (dotted curves) and the
combination of both methods (dashed curves).
The most obvious point to note is that for the smaller outer radius of
the annulus, the magnification method yields smaller errors in $q$
than the shear method, for all values of $a$ and $q$ (see lower left
panel), whereas for the error in the amplitude $a$, the relative
merits of both methods depend on $a$ and $q$. For the larger outer
aperture radius, the amplitude of the mass distribution is much better
determined from the shear method, and the relative accuracy of the slope
determination depends mainly on $q$: for steep profiles, $q\gtrsim 1$,
the shear method yields more accurate results, whereas the
magnification method is superior for flatter profiles. Except for the
shear method, the accuracy increases with decreasing $q$ and
increasing $a$, or in other words, with increasing lens strength. For
the smaller outer radius of the aperture, the flattest profiles yield
the largest errors for the shear method, which is counter-intuitive
and must be related to the accidental cancellations of effects in the
radial profile of the reduced shear, seen in Fig.\ts 1. 

\subs{6.2 Numerical simulations: comparison of the likelihood analysis
with $\chi^2$ statistics}
For each method, we now quantify the agreement between the
distribution of $2\Delta\ave{\ell} = \ave{\ell}(\pi) -
\ave{\ell}(\pi_\t)$ and that expected if $2\Delta\ave{\ell}$ followed
a perfect $\chi^2_2$ distribution. Ten thousand catalogs of lensed
galaxies were generated using a Family A lens model, with true
parameters $\pi_\t$ describing the cluster, and the log-likelihood
functions were minimized to obtain the best fitting parameters $\pi$
for each realization. For each $\pi$ we then calculate
$2\Delta\ave{\ell}$ and derive the cumulative probability distribution
$P(>2\Delta\ave{\ell})$, which can be compared with the perfect
distribution $P(>\chi^2_2)$. In Fig.\ts 5 the ratio of
$P(>2\Delta\ave{\ell}):P(>\chi^2_2$) is plotted against
$2\Delta\ave{\ell}$ for each of the magnification, shear and combined
methods. The deviation from a $\chi^2_2$ distribution is very small
for the magnification method, and slightly larger for the shear and
combined methods, but still very acceptable: the deviation from
$\chi^2_2$ statistics is such that the recovered values of $\pi$ for
about 94.4\% of the realizations lie within the 95.4\%-confidence
interval.

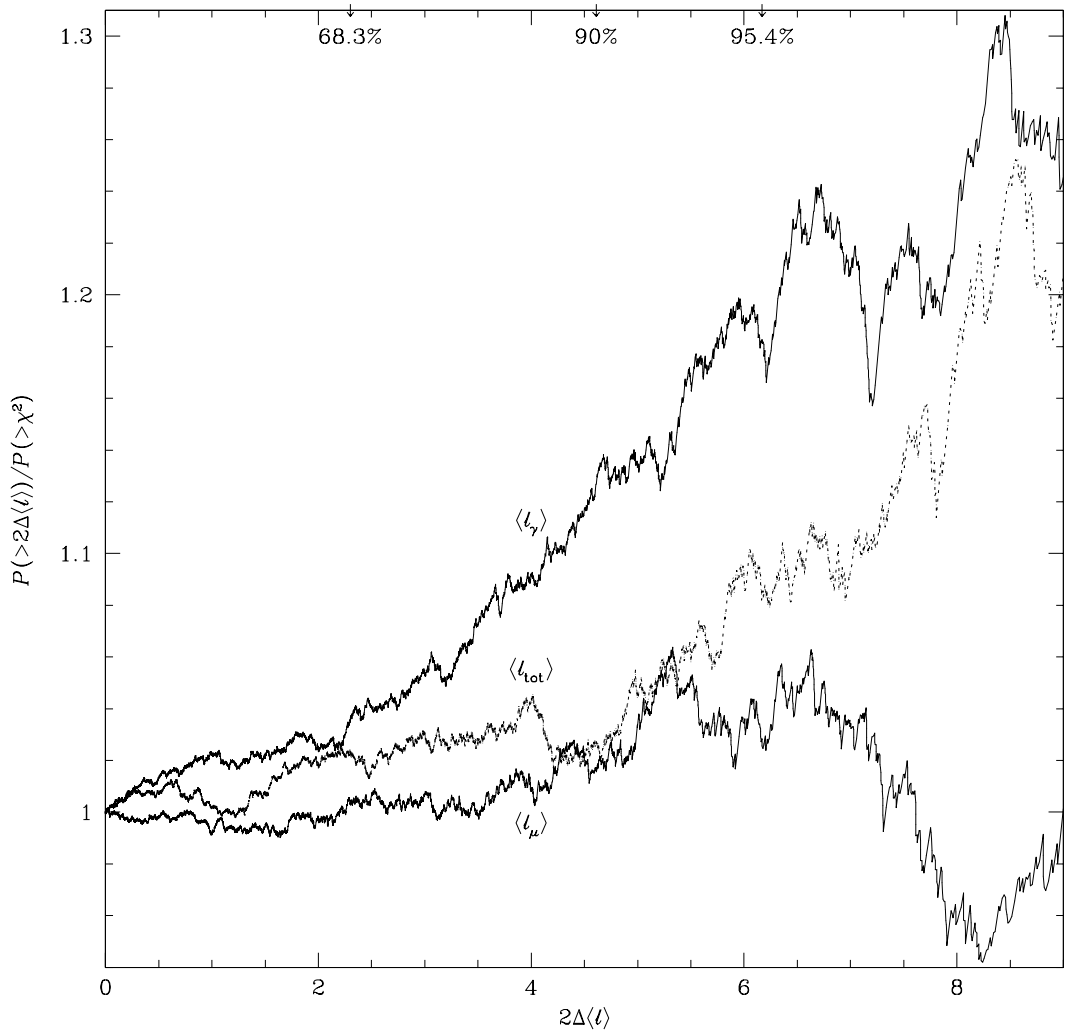
\xfigure{5}{For each of the magnification (lower curve), shear (upper
curve) and combined methods (dashed middle curve), we show the ratio
of $P(>2\Delta\ave{\ell})$ to that expected if $2\Delta\ave{\ell}$
followed a $\chi^2$ distribution versus $2\Delta\ave{\ell}$. On the
top edge of the plot the 68.3\%-, 90\%- and 95.4\%-confidence
intervals are marked}{fig6b.tps}{15}

The scatter of the recovered values of $\pi$ in the
$a$--$q$ plane is also consistent with the ensemble-averaged
log-likelihood contours, as we will demonstrate below. We can also ask
how well each of the methods can distinguish {\it between}
the Families of lens models. To these ends, we generated 500 
catalogs of lensed galaxies using the
prescription for numerical simulations given in Sect.\ts 5, with a 
Family A lens model to describe the cluster. The 
log-likelihood functions $\ell_\mu$,
$\ell_\gamma$ and $\ell_{\rm tot}$ were independently 
minimized under a Family A lens model and a Family B lens model, 
to obtain the best fitting parameters and their corresponding log-likelihood
values; we refer to these as $\ell_\mu$(A), $\ell_\gamma$(A) and
$\ell_{\rm tot}$(A) where the minimization is performed under a
Family A lens model, and $\ell_\mu$(B), $\ell_\gamma$(B) and 
$\ell_{\rm tot}$(B) where a Family B lens model is used during the
minimization. By comparing the log-likelihood values corresponding to
recovery using each of the two Families, we determined how frequently the
minimum corresponds to recovery by Family A. 

In the upper, middle and lower panels of Fig.\ts 6 we show our results for the
magnification, shear and combined methods respectively. The left-hand
panels show parameter recovery under a Family A lens model,
whereas a Family B lens model was used to recover the best fitting
parameters indicated on the right-hand panels. Contours of constant
average likelihood are marked on the left-hand panel; note that the
distributions of Monte Carlo points are consistent with these.

In 47\% of the realizations $\ell_\mu ({\rm A})< \ell_\mu$(B)
indicating that the magnification method cannot discriminate between
Family A and Family B.  The middle panel of Fig.\ts 6 shows that the
shear method performs slightly better: in 64\% of the realizations
$\ell_\gamma$(A)$<\ell_\gamma$(B). The results for recovery using the
combined method are shown in the bottom panel; $\ell_{\rm tot}$(A)$<
\ell_{\rm tot}$(B) in 63\% of the realizations.

We repeated the same procedure, generating catalogs with a Family B
lens model, and performing the minimization of $\ell_\mu$,
$\ell_\gamma$ and $\ell_{\rm tot}$ under a Family B lens model and
under a Family A lens model; our results are shown in Fig.\ts
7. Contours of constant average likelihood are marked on the left-hand
panels and the Monte Carlo points are consistent with these.

Again, the shear method (middle panel) fares a little better than the
magnification method (top panel) in discriminating
between the Families: in 57\% of the realizations,
$\ell_\mu$(B)$<\ell_\mu$(A) and in 73\% of the realizations 
$\ell_\gamma$(B)$<\ell_\gamma$(A). The combined method (bottom panel) gives
 $\ell_{\rm tot}$(B)$<\ell_{\rm tot}$(A) in 78\% of the realizations.

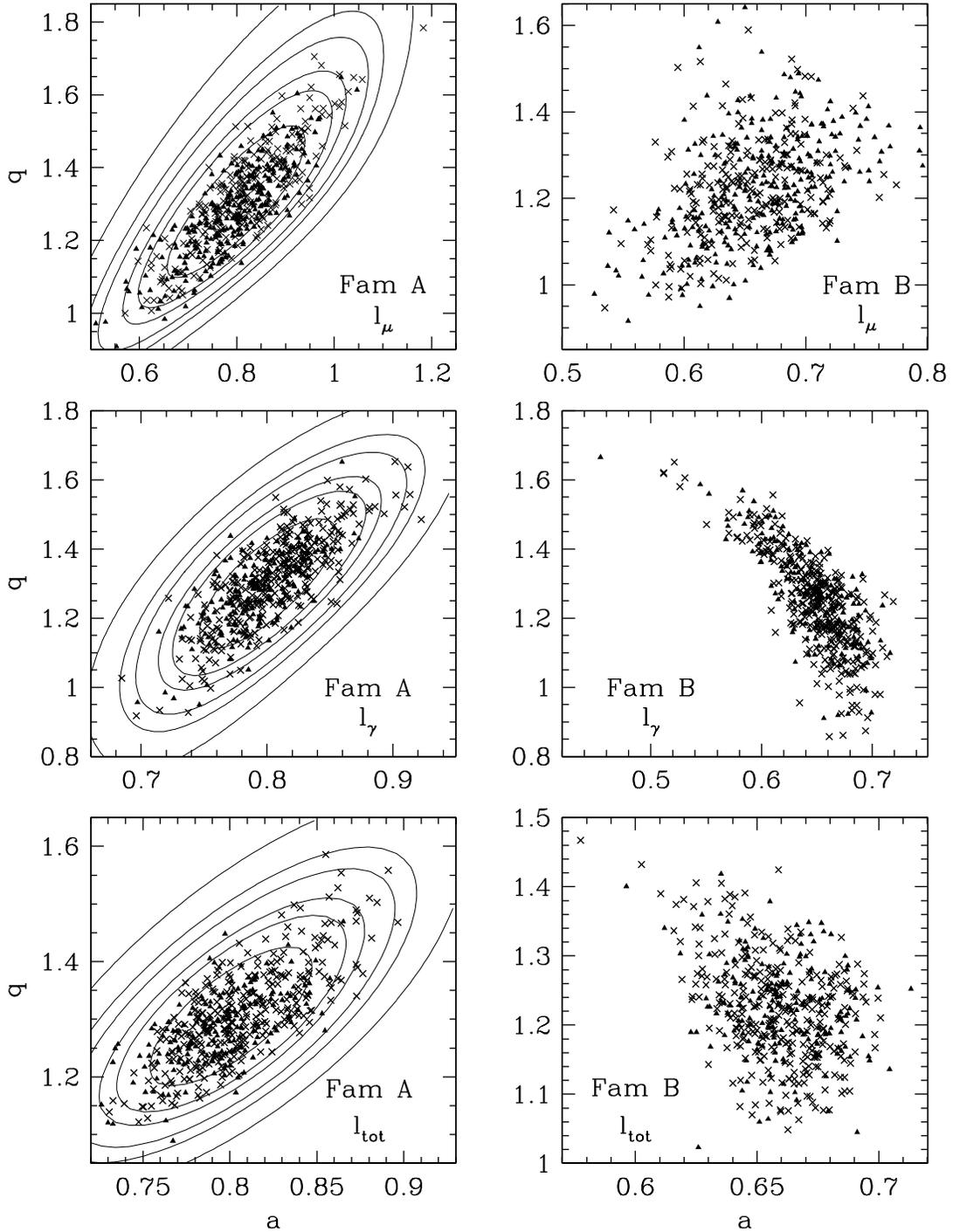
\xfigure{6}{A Family A input model ($a=0.8$, $q=1.3$) was used to
generate lensed galaxy catalogs. For each of the magnification (top
panel), shear (middle panel) and combined methods (bottom panel) the left-hand
(right-hand) panel shows parameter recovery under a Family A (Family
B) model and the crosses (triangles) indicate where the log-likelihood 
function is lowest when the minimization is performed using a Family A 
(Family B) lens model. On the left-hand panels we mark contours of 
constant $\Delta\ave{\ell}$ with levels the same as in Fig.\ts 2}{fig7.tps}{15}

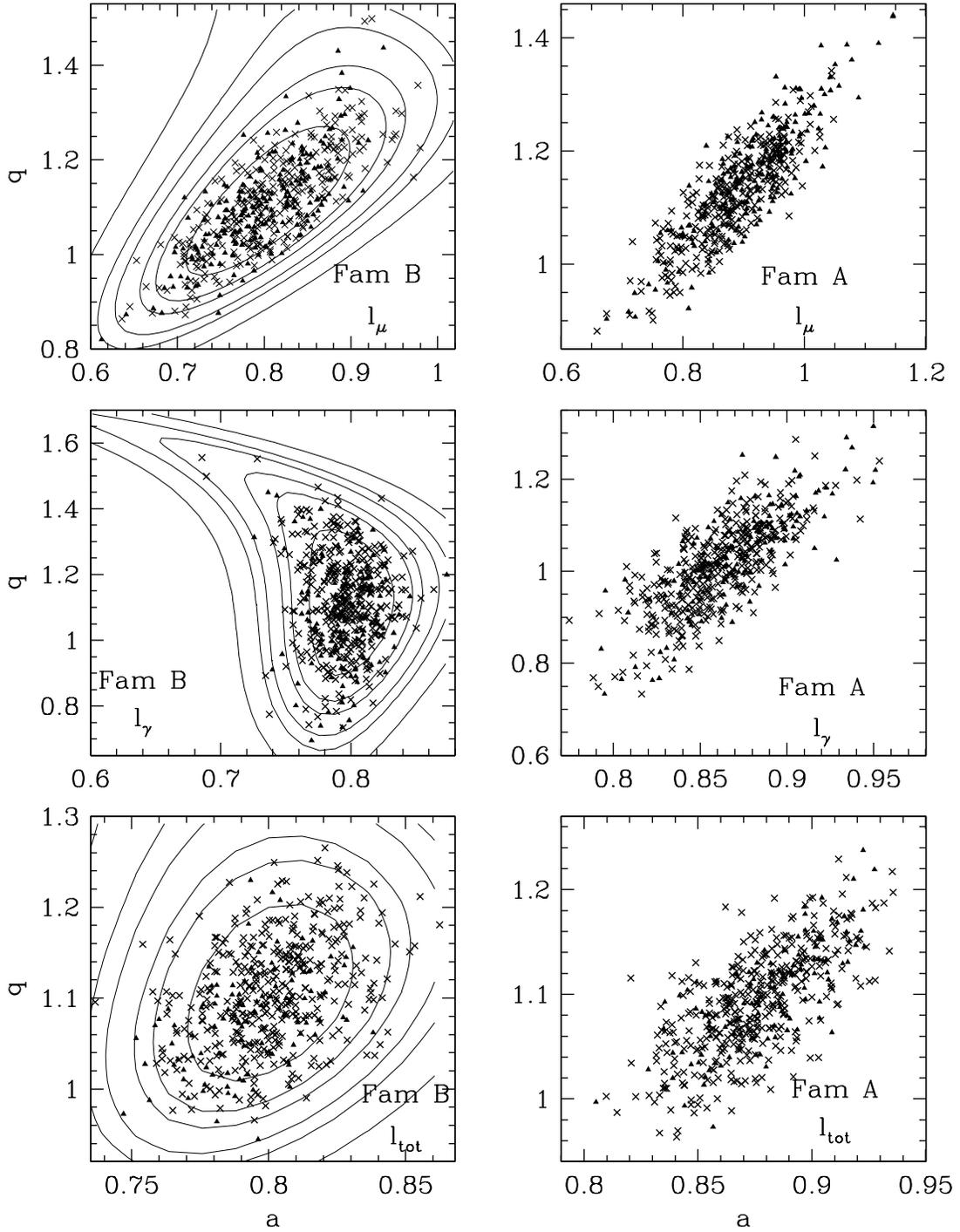
\xfigure{7}{A Family B input model ($a=0.8$, $q=1.1$) was used to
generate lensed galaxy catalogs. For each of the magnification (top
panel), shear (middle panel) and combined methods (bottom panel) the left-hand
(right-hand) panel shows parameter recovery under a Family B (Family
A) model and the crosses (triangles) indicate where the log-likelihood 
function is lowest when the minimization is performed using a Family B 
(Family A) model. On the left-hand panels we mark contours of constant
$\Delta\ave{\ell}$ with levels the same as in Fig.\ts 2}{fig8.tps}{15}

\subs{6.3 Uncertainties in the unlensed number counts}
Up to now we have assumed that the number density of background
galaxies at the flux threshold of the observations is known. In fact,
the shape of the likelihood contours in Figs.\ts 2 and 3 are such that
they closely follow the curve of constant expected total galaxy number
in the annulus, shown as the middle of the heavy dashed curves in
those figures which are curves of constant
$\ave{N_\mu}=\ave{N_\mu}_{\rm t}$.  This implies that to first order,
the magnification information is provided by this total number. As
indicated by the other two heavy solid curves in these figures, a
change of the number by $\pm 5\%$ changes the curve of constant
$\ave{N_\mu}$ quite drastically. Given that a high precision of the
number density of very faint galaxies may be difficult to
achieve, we shall now consider two different assumptions: first, we
shall assume that the unlensed number density is not determined from
any calibration frame, but is obtained from the present data set
itself. After that, we shall consider the (more realistic) assumption
that the unlensed number counts is known, but with some uncertainty.
To this end, we reinsert the density-dependent
term, which we discarded in (11), into the log-likelihood function,
which then reads
$$
\ell_\mu =n_\mu\,I
+(1-\beta)\sum_{i=1}^{N_\mu} \ln\mu(\vc\theta_i)-N_\mu\,\ln n_\mu\;,
\eqno (40)
$$
where we have defined
$$
I:=\int\d^2\theta\;[\mu(\vc\theta)]^{\beta-1} \;.
\eqno (41)
$$
This expression can now be minimized with respect to the unknown
density $n_\mu$, which yields, not unexpectedly,
$$
n_\mu={N_\mu\over I}\; .
$$
Inserting this value into (40) yields the new log-likelihood function,
up to a model-independent constant,
$$
\hat\ell_\mu=(1-\beta)\sum_{i=1}^{N_\mu} \ln \mu(\vc\theta_i)+N_\mu
\ln I\; .
\eqno (42)
$$
Following the techniques applied in Sect.\ts 4, the ensemble average
of $\hat\ell_\mu$ can be calculated,
$$\eqalign{
\ave{\hat\ell_\mu}&=\bar n_\mu\rund{\int\d^2\theta\;
[\mu_{\rm t}(\vc\theta)]^{\beta-1}}
\ln I \cr
&+\bar n_\mu(1-\beta)\int\d^2\theta\;
[\mu_{\rm t}(\vc\theta)]^{\beta-1}\,\ln \mu(\vc\theta)\; , \cr }
\eqno (43)
$$
where now $\bar n_\mu$ is the true number density of galaxies. By
differentiation with respect to $\pi_i$, one can easily show that
$\ave{\hat\ell_\mu}$ has a minimum when $\pi=\pi_{\rm t}$. 

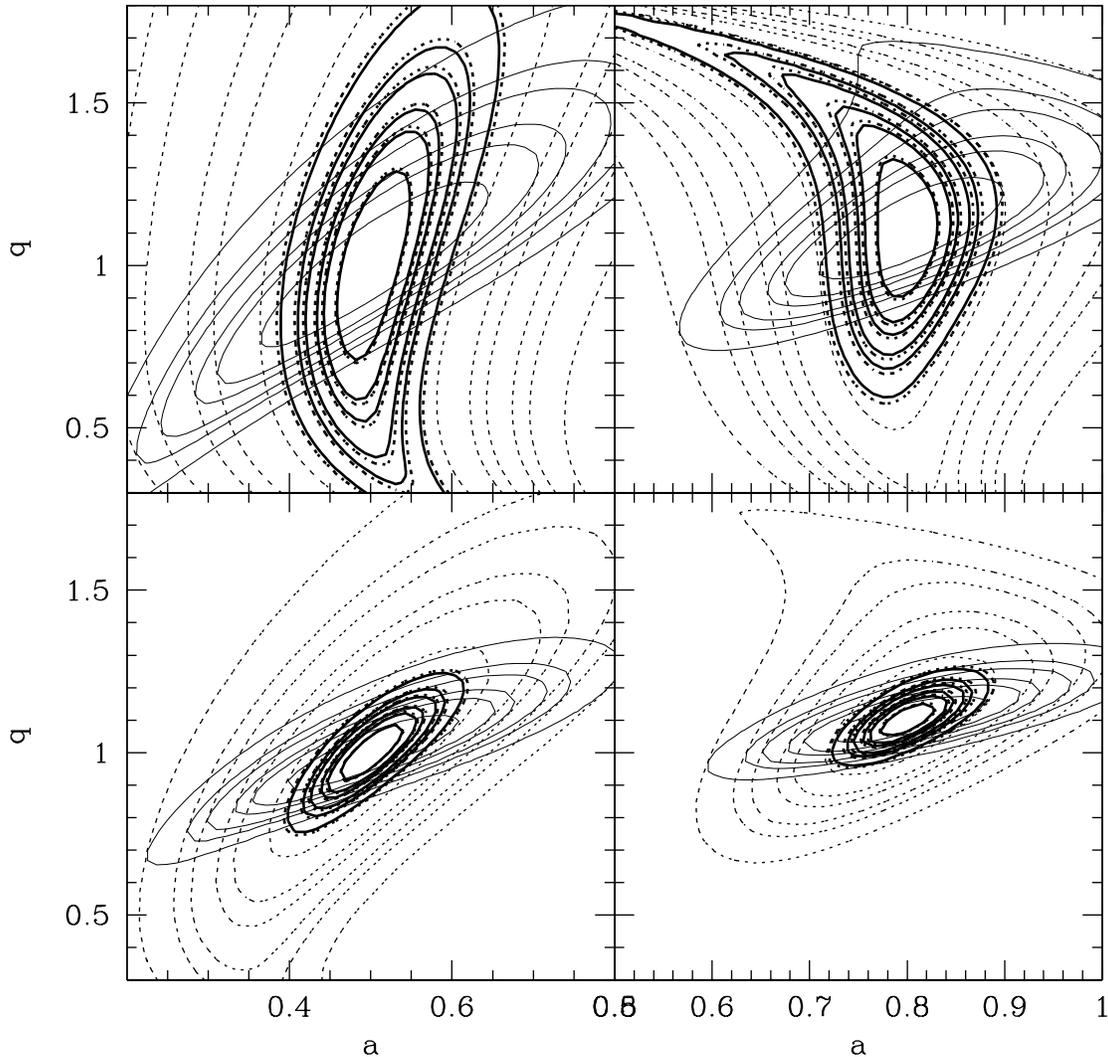
\xfigure{8}{Similarly to Figs.\ts 2 and 3, contours of constant
$\Delta\ave{\ell}$ are plotted; the contour levels are the same as
described in Fig.\ts 2. The inner radius of the annulus is
$\theta_{\rm in}=0\arcminf6$, the outer radius is $\theta_{\rm
out}=4'$ (upper panels) and $15'$ (lower panels); left (right) panels
correspond to models of Family A (Family B).
The thin dashed contour
correspond to $\ave{\hat\ell_\mu}$, the thin solid contours to
$\ave{\ell_\mu}$, the heavy dashed contours to $\ave{\ell_\gamma}$,
and the heavy solid contours to $\ave{\hat\ell_{\rm
tot}}:=\ave{\ell_\gamma+\hat\ell_\mu}$. The input model is $a=0.5$,
$q=1.0$ for Family A, and $a=0.8$, $q=1.1$ for Family B}{fig5.tps}{15} 

In Fig.\ts 8 we have plotted contours of constant likelihood of
$\ave{\hat\ell_\mu}$, for two values of the outer radius of the annulus,
together with the corresponding contours of $\ave{\ell_\mu}$,
$\ave{\ell_\gamma}$, and $\ave{\hat\ell_{\rm
tot}}:=\ave{\ell_\gamma+\hat\ell_\mu}$. From the figure it becomes
immediately clear that dropping the assumption about the knowledge of
the unlensed number density leads to a drastic loss of information from
the magnification method. The thin dashed contours in Fig.\ts 8 which
correspond to $\ave{\hat\ell_\mu}$ are much wider than the solid contours
corresponding to $\ave{\ell_\mu}$. In addition, whereas the contours
of $\ave{\ell_\mu}$ were seen to be substantially misaligned relative
to those of $\ave{\ell_\gamma}$, so that the combination of shear and
magnification information substantially decreases the error region in
parameter space relative to any single one of these methods, the
contours of $\ave{\hat\ell_\mu}$ are basically parallel to those of 
$\ave{\ell_\gamma}$ (shown as heavy dashed curves in Fig.\ts
8). Therefore, the combination of both methods, in the absence of
knowledge of the unlensed number density, yields only slightly more
accurate parameter estimates than the shear method alone.

To understand this point, one needs to realize why the magnification
information turned out to be so powerful when $n_\mu$ is known. As
we have seen in Fig.\ts 1, the reduced shear of models with different
$q$ is very similar, owing to the mass-sheet degeneracy discussed in
the introduction. Knowing $\bar n_\mu$, the mass-sheet degeneracy is
broken, which leads to the sensitivity of the magnification to the
slope of the density profile. However, if $\bar n_\mu$ is unknown, the
mass-sheet degeneracy remains, and the discussion of the relative
merits of shear and magnification methods in Sect.\ts 2 prevails. 

We now make the more realistic assumption that the unlensed number
density is known approximately. Suppose one knows the value of $\bar
n_\mu$ with a fractional accuracy of $\eta$; one can then supplement
the likelihood fucntion with a prior, taken to be a Gaussian in
$n_\mu$, with mean $\bar n_\mu$, and dispersion $\sigma_n=\eta \bar
n_\mu$. Hence, the corresponding log-likelihood function becomes
$$
\ell_\mu =n_\mu\,I
+(1-\beta)\sum_{i=1}^{N_\mu} \ln\mu(\vc\theta_i)-N_\mu\,\ln n_\mu
+\rund{ (n_\mu-\bar n_\mu)^2\over 2 (\eta \bar n_\mu)^2}
\;.
\eqno (44)
$$
As before, this can be minimized with respct to $n_\mu$, yielding
$$
{n_\mu \over \bar n_\mu}={1\over 2}\rund{1-\eta^2 \bar n_\mu I}
+\sqrt{{1\over 4}\rund{1-\eta^2 \bar n_\mu I}^2+\eta^2 N_\mu} \;.
\eqno (45)
$$
Inserting this value for $n_\mu$ into (44) yields the new
log-likelihood function, which we shall denote as
$\ell_\mu^{(\eta)}$. Taking the ensemble average in the present case
is slightly more difficult, owing to the occurrence of $N_\mu$ in the
square-root. However, if we neglect the Poisson-fluctuation in the
ensemble averaging in (17), which is a very good approximation when
$\ave{N_\mu}_{\rm t}$ is large, the ensemble average can be obtained as
$$
\ave{\ell_\mu^{(\eta)}}=\ave{n_\mu}I + \bar n_\mu (1-\beta)
\int\d^2\theta\;\mu_{\rm t}^{\beta-1}\,\ln \mu
-\ave{N_\mu}_{\rm t}\,\ln \ave{n_\mu} 
+ { (\ave{n_\mu}-\bar n_\mu)^2\over 2 (\eta \bar n_\mu)^2}\; ,
\eqno (46)
$$
where 
$$
{\ave{n_\mu} \over \bar n_\mu}={1\over 2}\rund{1-\eta^2 \bar n_\mu I}
+\sqrt{{1\over 4}\rund{1-\eta^2 \bar n_\mu I}^2+\eta^2
\ave{N_\mu}_{\rm t}} \;.
\eqno (47)
$$
Although this expression looks fairly complicated, it is easy to show
that $\ave{\ell_\mu^{(\eta)}}$ has a minimum when $\pi$ takes the value
of the true model. For $\eta\to 0$, this log-likelihood reduces, up to
additive constants, to the one for which the unlensed number density
was assumed to be known, i.e., eq.(11), whereas for large $\eta$, it
approaches (43). 

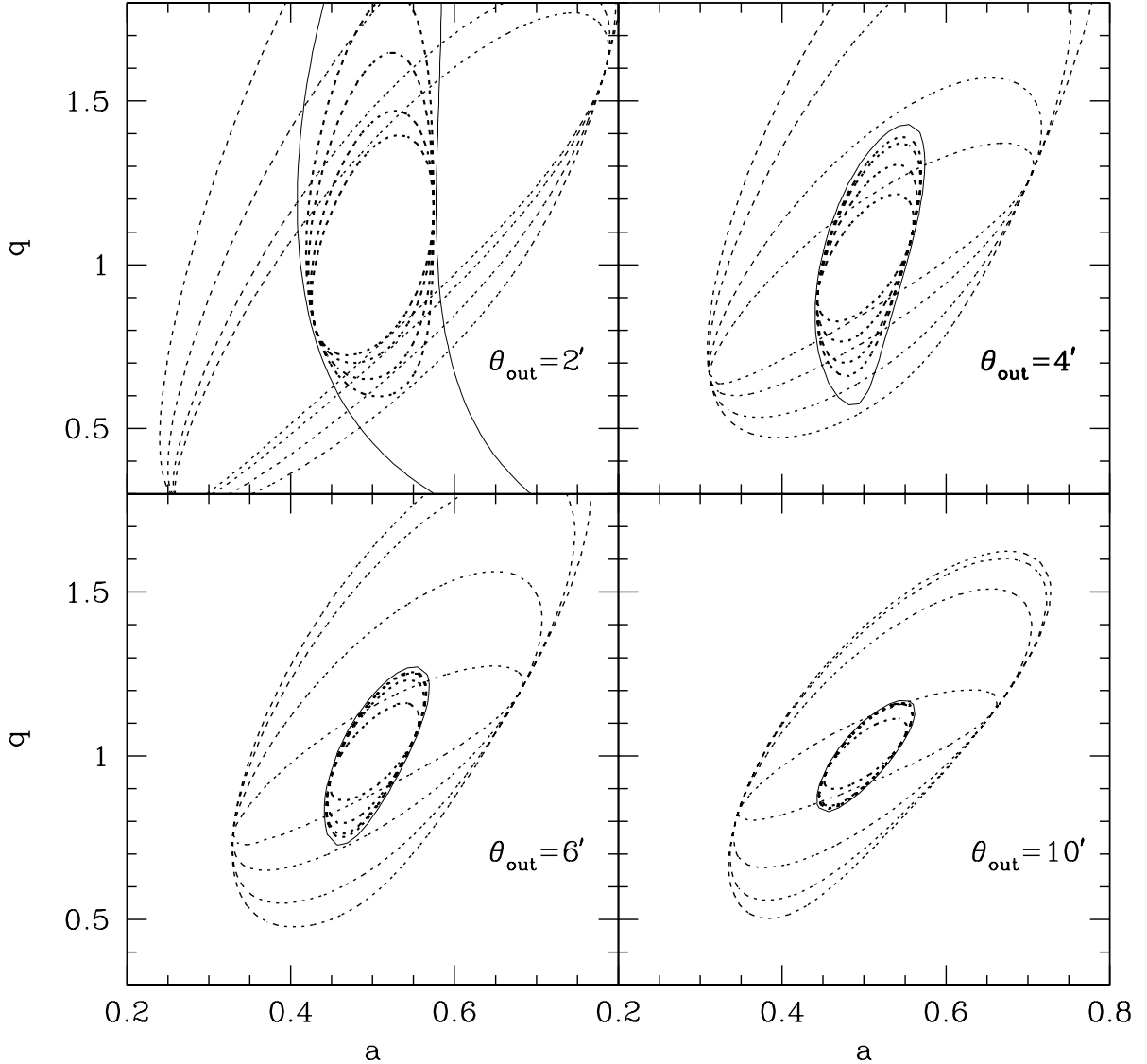
\xfigure{9}{For the case that the unlensed number counts are known
with a fractional accuracy $\eta$, the light dashed contours plot the
90\%-confidence contours of parameter estimates as obtained from the
ensemble-averaged log-likelihood function (46). The model $a=1/2$,
$q=1$ of Family A has been taken as input. The inner radius of the
annulus in which data are assumed to be is $\theta_{\rm in}=0\arcminf
6$, and the outer radius is different in each panel. The solid contour
is the 90\%-confidence contour as obtained from the shear, and the
heavy dashed curves from combining shear and magnification
constraints. The sets of four dashed curves are for $\eta=0$, 0.02,
0.04, and 0.06.}{fig10.tps}{16}

In Fig.\ts 9 we have plotted the 90\%-confidence contours as obtained
from $\ave{\ell_\mu^{(\eta)}}$, using models of Family A, for various
outer radii of the data annulus. Four values of the fractional
uncertainty $\eta$ in the number density have been chosen, ranging
from $\eta=0$ -- for which the contours agree with the corresponding
ones in Fig.\ts 2 -- to $\eta=0.06$. In addition, the 90\%-confidence
contour for the shear and those for the combined method are
plotted. As expected, by increasing $\eta$, the confidence region
increases, and the relative increase is larger, the larger
$\theta_{\rm out}$. The reason for this is that for the small annulus,
the relative change of the number density by magnification is much
larger than the fractional uncertainty, whereas this is no longer true
at larger radii. In fact, as can be seen from the lower right panel in
Fig.\ts 9, the confidence contours seem to approach an asymptotic
form quickly for large annuli, implying that the prior information
becomes irrelevant unless $\eta$ is very small, since the unlensed
number density is then more accurately determined from the data set
itself. Thus, the outer-most contour in the lower-right panel in
Fig.\ts 9 is very similar to the corresponding one in Fig.\ts 8,
where no prior information about the number density was assumed. Thus,
the larger the region from which data are used to determine the mass
profile parameters, the more accurate the unlensed number density
needs to be known, for a given accuracy relative to that obtainable
from the shear method. In particular, we conclude that the
magnification method considered here seems to be most powerful for data
sets which do not extend far from the cluster center, whereas for
wide-field imaging data, the shear method is superior.

Besides observational difficulties to determine the unlensed number
density with a given selection function from calibration data, the
anglar correlation function of galaxies provides a fundamental limit
down to which the unlensed number density can be determined within any
given region. To this end, we have calculated the fluctuations of the
number of galaxies in a circular aperture of radius $\theta$, using a
two-point correlation function of the form $\omega(\vt)=A
\vt^{-0.8}$. As a result, we find that $\delta N/N\propto
\theta^{-0.4}$, a slowly decreasing function of the
radius. Unfortunately, at the faint magnitude limits where the
magnification method is applied, no very accurate determination of the
angular two-point correlation function is available; the best
measurements has been carried out in the Hubble Deep Field (e.g.,
Villumsen et al.\ 1997, and references therein). Whereas at $B\sim 28$
a clear signal in the correlation function is seen on scales below
$\sim 1$\ts arcsec, the signal is too weak, and the solid angle of the
HDF too small to determine whether the power-law behaviour of $\omega$
extends beyond arcsecond scale. If one extrapolates the measurements
of Villumsen et al.\ (1997) to larger angular scales, then $\delta N/N
\sim 0.1$ on an angular scale of $\sim 4'$. In that case, the results
in Fig.\ts 9 suggest that the value of the magnification method is
rather limited.  More likely, though, the preceeding value is a vast
overestimate of the number count fluctuations due to angular
clustering at these magnitudes.

\sec{7 Discussion and conclusions}
Using maximum likelihood techniques, we have investigated the accuracy
with which weak gravitational lensing can recover the radial density
profile of clusters of galaxies. To keep the treatment as simple as
possible, we have assumed that the Einstein radius of the cluster is
known from strong lensing features, and that only the weak lensing
information from outside the Einstein radius is used; we also confined
the consideration to axially-symmetric mass distributions only.  We
have compared the shear method with the magnification method in form
of the number depletion. Although from a simple analytical
consideration the latter method is expected to yield less accurate
results, the mass sheet degeneracy causes the reduced shear to be
nearly degenerate along a one-dimensional subset of the two-parameter
mass models considered here. In some cases it then turns out that the
magnification method can constrain the parameters of the mass model
more accurately than the shear method. However, this is true only as
long as the unlensed number density of galaxy images is assumed to be
known quite accurately. In the absence of this knowledge, the
magnification method is inferior to the shear method. Since the number
density of galaxies obtained from `blank field' imaging depends not
only on the flux threshold (which requires accurate photometry), but
also on the seeing and sky brightness, this accurate number density
calibration may be difficult to achieve. The advantage of the shear
method is that no external calibration is required (the external
calibration is replaced by the assumption of random intrinsic
orientations). In addition, the angular correlation of faint galaxies
may present a fundamental obstacle for an accurate determination of
the unlensed counts in a given area, but too little is known about the
angular correlation function at flux limits at which the magnification
method is applied to make quantitative estimates of this effect at
present. 

Without accurate external calibration, our results show that large
areas around the cluster need to be mapped, since the external
calibration is effectively replaced by calibration from the outer
region of the data field. In order for this calibration to be accurate
the image quality needs to be homogeneous over the wide field. Of
course, since the measurement of shear requires superb observing
conditions, in particular with regards to the seeing, the number count
magnification method may be applied to photometric data set which
cannot be used for the shear method.

Although we have not investigated the magnification effect based on
the change of image size at constant surface brightness (Bartelmann \&
Narayan 1995), the previous remarks on external calibration apply
there as well.

External calibration may be possible using HST observations where the
observing conditions are much more stable than from the ground. Hence,
the claim by Broadhurst (1999) that the mass profile of the cluster
Cl0024+17 can be better constrained with the magnification method than
with the shear method is not contradicted by our results here; in fact,
if the external calibration of the number density is possible for
these HST images, the results shown in Figs.\ts 2 and 3 indicates a
higher accuracy of the shape parameter than possible with the shear
method. 

We would like to thank Matthias Bartelmann and Tom Broadhurst for
several fruitful discussions, and also the former for carefully
reading the manuscript.  This work was supported by the TMR Network
``Gravitational Lensing: New Constraints on Cosmology and the
Distribution of Dark Matter'' of the EC under contract
No. ERBFMRX-CT97-0172 and the ``Sonderforschungsbereich 375-95 f\"ur
Astro--Teil\-chen\-phy\-sik" der Deutschen
For\-schungs\-ge\-mein\-schaft.

\sec{Appendix}
We derive here the dispersion $\sigma$ of the image ellipticity
distribution. As in (2) let $p_\eps(\eps^\s)$ be the probability
density of the intrinsic source ellipticities. Without loss of
generality, we choose the reduced shear $g$ to be real, and write
$\eps^\s=x{\rm e}^{{\rm i}\vp}$. Then, the transformation (1) reads
$$\eqalign{
\eps_1&=g+(1-g^2)\,x\,{gx+\cos\vp\over 1+2gx\cos\vp+g^2x^2}\;,\cr
\eps_2&=(1-g^2)\,x\,{\sin\vp\over 1+2gx\cos\vp+g^2x^2}\;.\cr }
\eqno (A1)
$$
Then, the dispersion of the observed ellipticities in the two
directions becomes
$$\eqalign{
\sigma_\Vert^2&=\int_0^1\d
x\;x\,p(x)\int_0^{2\pi}\d\vp\;\rund{\eps_1-g}^2 \cr
&=(1-g^2)^2\int_0^1\d x\;x^3\,p_\eps(x)
\int_0^{2\pi}\d\vp\;\rund{gx +\cos\vp\over 1+2gx\cos\vp+g^2x^2}^2  \cr
&=(1-g^2)^2\pi\int_0^1\d x\; {x^3 p_\eps(x)\over 1-x^2 g^2}
=\sigma_\bot^2 \;. \cr }
\eqno (A2)
$$

\def\ref#1{\vskip1pt\noindent\hangindent=40pt\hangafter=1 {#1}\par}

\sec{References}
\ref{Bartelmann, M. \& Narayan, R.\ 1995, ApJ 451, 60}
\ref{Bartelmann, M., Narayan, R., Seitz, S., Schneider, P.\ 1996,
ApJ 464, L115}
\ref{Bonnet, H., Mellier, Y. \& Fort, B.\ 1994, ApJ 427, L83}
\ref{Broadhurst, T.J., Taylor, A.N. \& Peacock, J.A.\ 1995, ApJ 438,
49}
\ref{Broadhurst, T.J.\ 1999, talk given at the VLT Opening Symposium,
March 1--4, 1999, Antofagasta, Chile}
\ref{Canizares, C.R.\ 1992, ApJ 263, 508}
\ref{Clowe, D., Luppino, G.A., Kaiser, N., Henry, J.P. \& Gioia, I.\ 1998,
ApJ 497, L61}
\ref{Fahlman, G., Kaiser, N., Squires, G. \& Woods, D.\ 1994, ApJ 437,
56}
\ref{Falco, E.E., Gorenstein, M.V., Shapiro, I.I.\ 1985, ApJ 289, 1L}
\ref{Fischer, P.\ 1999, AJ 117, 2024}
\ref{Fort, B., Mellier, Y. \& Dantel-Fort, M.\ 1997, A\&A 321, 353}
\ref{Geiger, B. \& Schneider, P.\ 1998, MNRAS 295, 497}
\ref{Geiger, B. \& Schneider, P.\ 1999, MNRAS 302, 118}
\ref{Hoekstra, H., Franx, M., Kuijken, K., Squires, G.\ 1998, ApJ 504,
636}
\ref{Kaiser, N. \& Squires, G.\ 1993, ApJ 404, 441}
\ref{Kaiser, N., Squires, G., Fahlmann, G. G., Woods, D.\ 1994,
preprint CITA-94-40}
\ref{Kaiser, N.\ 1995, ApJ 439, L1}
\ref{Kaiser, N., Squires, G., Broadhurst, T.\ 1995, (KSB95), ApJ 449, 460}
\ref{Kaiser, N., Wilson, G., Luppino, G., Kofman, L., Gioia, I.,
Metzger, M., Dahle, H. \ astro-ph/9809268, submitted to ApJ}
\ref{Kochanek, C.S.\ 1990, MNRAS 247, 135}
\ref{Lombardi, M. \& Bertin, G.\ 1998a, A\&A 335, 1}
\ref{Lombardi, M. \& Bertin, G.\ 1998b, A\&A 330, 791}
\ref{Luppino, G. A., Kaiser, N.\ 1997, ApJ 475, 20}
\ref{Press, W.H., Teukolsky, S.A., Vetterling, W.T. \& Flannery, B.P.\
1992, in Numerical recipes in FORTRAN. The art of scientific
computing, Cambridge: University Press, 2nd Edn.}
\ref{Schneider, P.\ 1984, A\&A 140, 119}
\ref{Schneider, P., Ehlers, J. \& Falco, E.E.\ 1992, {\it Gravitational
lenses}, Springer: New York}
\ref{Schneider, P.\ 1995, A\&A 302, 639}
\ref{Schneider, P., Seitz, C.\ 1995, A\&A 294, 411}
\ref{Schramm, T. \& Kayser, R.\ 1995, A\&A 289, 5L}
\ref{Seitz, C. \& Schneider, P.\ 1995, A\&A 297, 287}
\ref{Seitz, C., Kneib, J.-P., Schneider, P. \& Seitz, S.\ 1996, A\&A
314, 707}
\ref{Seitz, S. \& Schneider, P.\ 1996, A\&A 305, 383}
\ref{Seitz, C. \& Schneider, P.\ 1997, A\&A 318, 687}
\ref{Seitz, S., Schneider, P., Bartelmann, M.\ 1998, A\&A 337, 325}
\ref{Squires, G. \& Kaiser, N.\ 1996, ApJ 473, 65}
\ref{Squires, G. et al.\ 1996a, ApJ 461, 572}
\ref{Squires, G., Kaiser, N., Fahlman, G., Babul, A. \& Woods, D.\ 1996b,
ApJ 469, 73}
\ref{Taylor, A.N., Dye, S., Broadhurst, T.J., Benitez, N. \& Van
Kampen, E.\ 1998, ApJ 501, 539}
\ref{Tyson, J.A., Valdes, F. \& Wenk, R.A.\ 1990, ApJ 349, L1}
\ref{Villumsen, J.V., Freudling, W. \& Da Costa, L.N.\ 1997, ApJ 481, 578}
\ref{Webster,R.L.\ 1985, MNRAS 213, 871}

\end